\documentclass[12pt,preprint]{aastex}

\shorttitle{Dust in Compact PNe}
\shortauthors{Lee \& Kwok}

\begin{document}

\title{Dust Extinction in Compact Planetary Nebulae}

\author{T.-H. Lee\altaffilmark{1}}
\affil{Department of Physics and Astronomy, University of
  Calgary, Calgary, AB T2N 1N4, Canada}
\email{thlee@noao.edu}

\and

\author{S. Kwok}
\affil{Institute of Astronomy \& Astrophysics, Academia Sinica, Taipei
  106, Taiwan\\ and\\Department of Physics and Astronomy, University of
  Calgary, Calgary, AB T2N 1N4, Canada }
\email{kwok@asiaa.sinica.edu.tw}

\altaffiltext{1}{Current address: National Optical Astronomy Observatories, Tucson, AZ 85719}

\begin{abstract}

The effects of dust extinction on the departure from axi-symmetry in
the morphology of planetary nebulae (PNs) are investigated through a
comparison of the radio free-free emission and hydrogen recombination
line images.  The dust exctinction maps of five compact PNs are
derived using high-resolution ($\sim$0\farcs1) H$\alpha$ and radio
images maps of the {\it HST} and {\it VLA}.  These extinction maps
are then analyzed by an ellipsoidal shell ionization model including
the effects of dust extinction to infer the nebulae's intrinsic
structure and orientation in the sky.  This method provides a
quantitative analysis of the morphological structure of PNs and
represents a step beyond qualitative classification of morphological
types of PNs.

\end{abstract}


\section{Introduction}


  In general, PNs have axisymmetric structures.  The classification of
  PNs into three basic types, round, elliptical and bipolar \citep{bal87}, is based
  on their axisymmetric appearances.  However, many PNs also possess
  non-axisymmetric structures.  \cite{sokerhadar02} proposed a scheme
  to classify PNs according to their departure from axisymmetry.  They
  considered six categories: (1) PNs where the central star is not at
  the center of the nebula; (2) PNs having one side brighter than the
  other; (3) PNs having unequal size or shape of the two sides; (4)
  PNs where the symmetry axis is bent, e.g., the two lobes in a
  bipolar PN are bent toward the same side; (5) PNs where the main
  departure from axisymmetry is in the outer regions, e.g., an outer
  arc; (6) PNs showing no large-scale departure from axisymmetry.
  They found that about 50\% of the analyzed sample has large-scale
  departures from axisymmetry.

  The problem of 
  departure from axisymmetry in PNs is usually investigated through the examination of optical images.  This poses a problem because
  dust also plays a role in making such asymmetry.  For example, in
  the sample analyzed by \cite{sokerhadar02}, more than half of the
  PNs they found to have departure from axisymmetry are the types with unequal
  intensity or size.  This kind of asymmetric appearance can also be
  caused by dust extinction, even if the intrinsic brightness of the
  nebula is axisymmetric.  It is very difficult to draw conclusions
  about the real shapes of PNs based only on their apparent optical
  forms.  In order to remove the bias caused by dust extinction while
  investigating the shapes of PNs, it is necessary to understand the
  dust distribution in PNs.

  Asymptotic Giant Branch (AGB) stars eject large amount of dust and gas in the form of stellar winds.  These stellar grains is a major source of dust grains
  in the Galaxy \citep{gehrz89}.  Although the details of dust
  formation are still poorly understood, the properties of circumstellar dust have been extensively studied in the
  infrared, e.g., by the 
  {\it Infrared Astronomy Satellite} ({\it IRAS}) and {\it Infrared Space Observatory} ({\it ISO}) missions.  Since PNs descend from AGB stars, the remnants of
  the AGB dust envelopes should still be present in PNs \citep{kwo82}.
The presence of dust in PNs was first detected in NGC~7027 via its
  infrared emission \citep{gillett67}.  
This infrared excess peaks around 30 $\mu$m and has a color temperature of $\sim$100~K.  
Although the existence of dust in PNs first came as a surprise, its widespread
  presence is now confirmed by  {\it IRAS} observations \citep{pottaschetal84, zhang91}.

  While the presence of dust in PNs is well established by infrared spectroscopy,  very little is known  about its spatial distribution.  Recent development in mid-infrared cameras has just begun to explore the dust distribution of PNs \citep[see, e.g.,][]{volk03, su04}.
However, a full mapping of the dust distribution of PNs has to wait until the development of large-format  mid-infrared cameras with high resolution and high sensitivity.  
An alternate and underutilized
  method to derive the dust distribution is by comparing the hydrogen
  recombination line map of a PN with its corresponding radio
  free-free continuum map.  The recombination line (e.g.~H$\alpha$)
  and free-free continuum emission are independent tracers of the
  distribution of the ionized gas, and both of their fluxes are
  proportional to $n_e^2 V$ under optically thin conditions, where
  $n_e$ is the electron density and {\it V} is the volume of the
  nebula.  If dust is present, 
they will have no effect on the long
  wavelength radio emission, but the  H$\alpha$ emission will be attenuated by dust.  Therefore
  the dust distribution can be indirectly obtained by comparing the
  H$\alpha$ and radio maps, assuming the dust is associated with the
  ionized gas.  This method has been applied to different objects by
  various authors \citep[e.g.,][]{vazquez99,eyres01}.  In this paper, we have derived
  optical depth maps of five compact PNs obtained with the same
  technique.  Compact PNs are selected for this study because their
  small angular sizes allow the assumption that any foreground
  interstellar extinction is homogeneous over the entire object, so
  that the different degrees of extinction over a PN are only related
  to its internal dust distribution.  

\section{Derivation of the optical depth distribution}
\label{sec:ionized}

In a fully ionized region, both the free-free emission and the recombination fluxes are proportional to $n_e^2 V$, where $n_e$ is the electron density.  The ratio of the free-free flux to the H$\beta$ recombination line fluxes under Case B (where recombination to the ground state is excluded) is given by the following equation \citep{pottasch84}:

\begin{equation}
\label{eq:ff-hb}
  \frac{S_{\nu}({\rm ff})}{F({\rm H}\beta)} = 2.51 \times10^{7}
  T_{e}^{0.53}\nu^{-0.1} Y ~~~\rm \frac{Jansky}{erg~cm^{-2} s^{-1}}~,
\end{equation}
  where $T_e$ is the electron temperature, and
\begin{equation}
  Y = 1 + \frac{n(He^{+})}{n_{p}} + 3.7\frac{n(He^{++})}{n_{p}} 
\end{equation}
  corrects for the contribution to the free-free flux from He ions.  For a He to H number ratio of 0.11 and for He to be equally divided between singly and doubly ionized forms, $Y$ has a value of 1.258.  Under Case B, the effective recombination coefficients for H$\alpha$ and H$\beta$ have a ratio of 2.85 \citep{hummer87}.  For $T_e=10^4$ K, eq.~\ref{eq:ff-hb} can be written as

\begin{equation}
\label{eq:depth}
  F_{{\rm exp}}({\rm H}\alpha) = 6.85 \times 10^{-10} (\nu/{\rm GHz})^{0.1}
  (S_{\nu}/{\rm Jy}) ~~~\rm erg~cm^{-2}~s^{-1} ~,
\end{equation}
  which can be used to calculate the expected flux of H$\alpha$
  emission without dust extinction from the flux density of radio
  continuum emission.

In order to derive the extinction map, we begin with the radio continuum and H$\alpha$ maps of the same PN at similar angular resolutions.
  The expected H$\alpha$ emission at each pixel is derived from the
  radio continuum map, and then divided by the observed H$\alpha$
  emission map for each object to obtain an extinction map, in which
  the dust optical depth at each pixel can be calculated.  The
  intensity emitted by an object and received by the observer is
  related by the optical depth $\tau$ as

\begin{equation}
\label{eq:absorption}
  F_{\rm obs} = F_{\rm exp} e^{-\tau}~,
\end{equation}
  where $F_{\rm exp}$ is the intensity emitted by the object, i.e., the
  intensity expected to be received by the observer if no dust is
  present.  Therefore the dust optical depth for H$\alpha$ emission is
  obtained by taking the natural logarithm of the ratio of the
  expected and observed H$\alpha$ flux:

\begin{equation}
\label{eq:tau}
  \tau_{{\rm H}\alpha} = \ln \frac{F_{exp}({\rm
  H}\alpha)}{F_{obs}({\rm H}\alpha)}~.
\end{equation}

\section{Sample selection and data processing}
\label{sec:data-dust}

\begin{figure}
\begin{center}
\includegraphics[width=\textwidth]{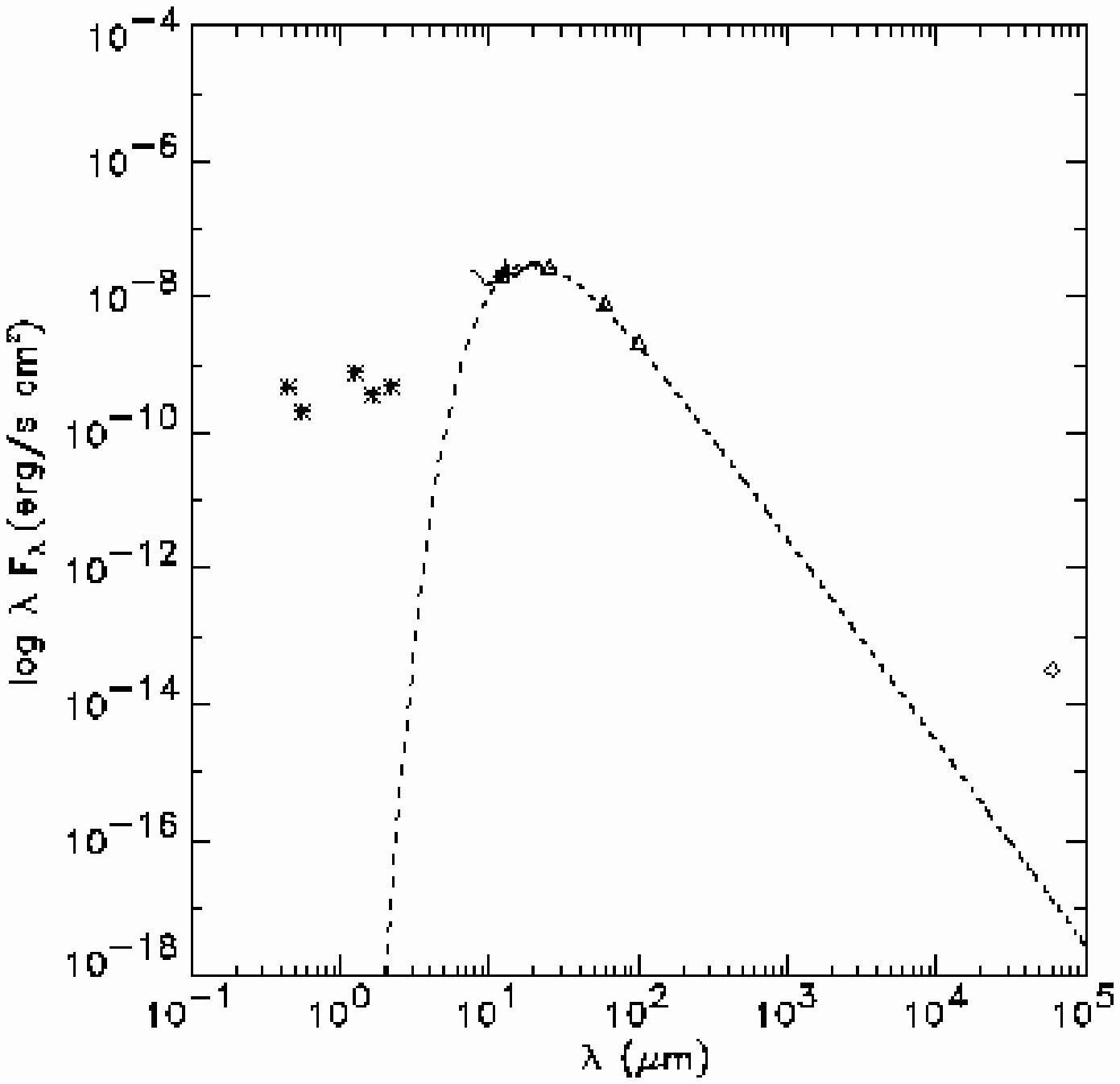}
\caption{The SED of BD+30$^\circ$3639, IC 5117, M~1-61, M~2-43, and M~3-45.
        The asterisk are ground-based optical and 2MASS measurements,
        the triangles are {\it IRAS} measurements, the crosses are
        {\it MSX} measurements, diamond are radio measurements, and
        the solid line is the {\it IRAS} LRS spectrum.  The dashed
        line is the fitted blackbody curve for dust emission.  The dust temperatures are
        BD+30$^\circ$3639: $T_{d} = 190 \rm K$, IC~5117: 180 K, M~1-61: 170 K, M~2-43: 200 K, and M~3-35: 180 K.} 
\label{sed}
\end{center}
\end{figure}

In this paper, we have selected five compact PNs with known high infrared excesses.  
 Figures \ref{sed} show the spectral energy distributions (SEDs) of these five PNs from optical to radio wavelengths, illustrating the   IR excess caused by dust.  The ground-based optical data are from \cite{acker92} and  \cite{acker81}.  The near-infrared {\it J}, {\it H}, and {\it
  K}-band photometry are from the Two Micron All Sky Survey (2MASS),
  extracted from the 2MASS All-Sky Catalog of Point Sources
  \citep{cutri03}.  The mid-infrared photometry includes {\it IRAS}
  photometry, {\it IRAS} low resolution spectra (LRS), and recent
  measurements from the {\it Midcourse Space Experiment} ({\it MSX}).
  The {\it MSX} measurements were extracted from the {\it MSX} Point
  Source Catalog \citep{egan03}.  Radio photometric measurements (mostly at $\lambda$ 2 and 6 cm) are from  \cite{aaquist90}, \cite{basart87}, and this study.
As  clearly shown in Fig.~\ref{sed}, the infrared excess of the PNs is well represented by
  a single blackbody curve with a temperature $\sim$ 200~K.  
 
\placefigure{sed}

High resolution H$\alpha$ and radio images of these PNs are available  
from the {\it HST} and the {\it VLA} respectively.  The basic information for
  the five objects is listed in Table \ref{tb:objects-dust}.  With these high resolution images, the point-to-point spatial extinction distribution in PNs can
  be obtained with angular resolutions as high as $\sim$ 0.1 arcsec.

\placetable{tb:objects-dust}

  For IC~5117, M~1-61, M~2-43, and M~3-35, the {\it HST} WFPC2
  H$\alpha$ images were obtained under the GO program 8307
  \citep{kwoksu03}.  
For BD+30$^\circ$3639, the {\it HST} WFPC2 H$\alpha$ image was first presented in \cite{harrington97} and the calibrated
  images were obtained from the {\it HST} archive.  The only image
  processing needed was to remove cosmic rays with the task CRREJ in
  IRAF.  The angular resolution FWHM of these images ranges from
  0\farcs085 to 0\farcs11.
 

  For BD+30$^\circ$3639, the $\lambda$6 cm radio image was from
  \cite{bryce97}.  It was obtained by combining VLA and Multi-Element
  Radio Linked Interferometer Network (MERLIN) observations, resulting
  in a very high angular resolution of $\sim$ 0\farcs08.  For the
  remaining four objects, the radio continuum maps were observed by
  Aaquist \& Kwok with the VLA at $\lambda$2 cm in A-configuration in
  March 1998.  The unpublished data were retrieved from the VLA
  archive and reduced using the package AIPS following the standard
  procedures of continuum data reduction, as summarized in Appendix A
  of the AIPS Cookbook \citep{nrao03}.
  After the data were calibrated, the final CLEAN
  images were generated with robust weighting in order to obtain
  images with minimum-noise levels while maintaining the highest
  angular resolution.  This ranges from $\sim 0\farcs12$ to $\sim 0\farcs15$
  in the final CLEAN maps.

\placetable{tb:fwhm}

  The operations taken to combine the radio and optical data and to
  derive the optical depth map are performed in the following sequence.  
  The {\it HST} images were first oriented
  north-up with Karma software.  Subsequent regridding and other steps
  were performed in AIPS.  Because the coordinates of the {\it HST}
  images have a uncertainty of $\sim 0\farcs5$, there is usually an
  offset in the positions of the same object in the radio and optical
  images.  To correct this, the H$\alpha$ images were shifted and
  aligned by eye to match with the radio continuum maps using features
  common in both images.  Table \ref{tb:shift} gives the offset of
  the coordinates for each object; a positive sign indicates the
  optical image has been shifted eastward or northward.  Each object's
  H$\alpha$ image was convolved with an extended Gaussian to match the
  beam size of the corresponding radio image, except for
  BD+30$^\circ$3639.  Because the radio continuum image of
  BD+30$^\circ$3639 has a higher resolution than its H$\alpha$ image,
  it was convolved with an extended Gaussian to match the FWHM of the
  H$\alpha$ image.  The radio image was transformed so that its pixel
  coordinate grid was consistent with that of the {\it HST} image.
  Division and natural logarithm operations were then performed to
  obtain an optical depth map as in eq.~\ref{eq:tau}.

\placetable{tb:shift}

\section{Results and Analysis}
\label{sec:results-dust}

\begin{figure}
\begin{center}
\includegraphics[width=\textwidth]{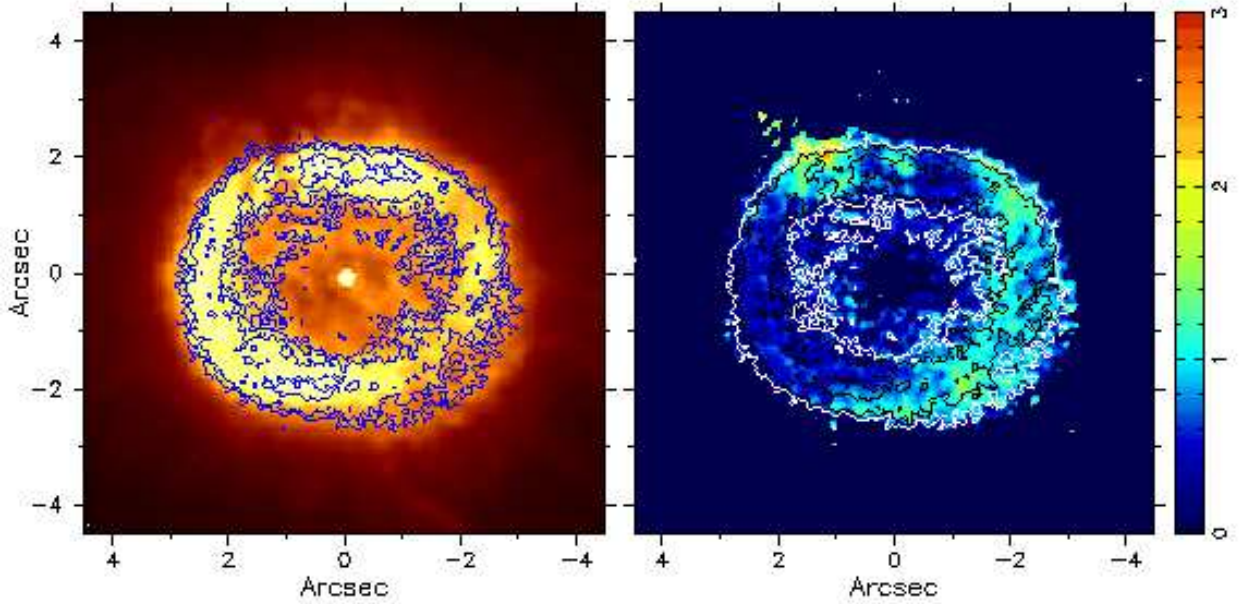}
\caption {Images (from top to bottom) of BD+30$^\circ$3639, IC~5117, M~1-61, M~2-43, M~3-35.
	Left panel: $\lambda$6 cm radio continuum (contours) on {\it HST}
	H$\alpha$ image (color, in square-root intensity scale).  Right panel: $\lambda$6 cm radio continuum
	(contours) on extinction map (color).  The color bar indicates
	$\tau$ values.  BD+30$^\circ$3639: The contour levels are $\sigma \times$ (3, 5,
	10, 15) with the noise level $\sigma = 4.5 \times 10^{-5}$
	Jy/beam.  IC~5117: the contour levels are $\sigma \times$ (3, 5, 10, 15,
	20, 25, 30, 40, 50) with the noise level $\sigma = 8.89 \times
	10^{-5}$ Jy/beam.   M~1-61: the contour
	levels are $\sigma \times$ (3, 5, 10, 15, 20, 25, 30, 35, 40,
	45) with the noise level $\sigma = 8.2 \times 10^{-5}$
	Jy/beam.  M~2-43: the contours
	levels are $\sigma \times$ (3, 5, 10, 20, 30, 40, 50, 60, 70,
	80) with the noise level $\sigma = 1.11 \times 10^{-4}$
	Jy/beam.  M~3-35: the color bar indicates $\tau$ values.  The contour
 	levels are $\sigma \times$ (3, 5, 10, 15, 20, 25, 30, 40, 50)
 	with the noise level $\sigma = 7.58 \times 10^{-5}$ Jy/beam.}
\label{depth}
\end{center}
\end{figure}

  Extinction maps showing the dust optical depth for the five compact
  PNs are presented in Fig.~\ref{depth}.  For comparison, the H$\alpha$   brightness and extinction maps for each object are shown side by  side, and the corresponding radio continuum contours are superposed
  on both maps.  All images are oriented with north up and east to the left.

\subsection{BD+30$^\circ$3639}

  The radio continuum image of BD+30$^\circ$3639 shows a ring-like
  structure with two peaks of roughly the same brightness in the north
  and south regions.  The H$\alpha$ image also shows a ring-like
  structure with two peaks.  However, the ring shows unequal
  brightness in the east and west regions, which indicates the ring
  suffers more dust extinction in the west as shown in the optical
  depth map.  The two high-extinction regions in the northeast and
  southwest are in roughly the same positions as the pair of
  high-velocity molecular knots found with CO line imaging by
  \cite{bachiller00}.

\placefigure{depth}

\subsection{IC~5117}

  The radio continuum image of IC~5117 shows a shell structure with
  two emission peaks.  Such a shape can be simply explained as a
  prolate ellipsoidal shell projected onto the plane of the sky
  \citep{aaquist91}.  The H$\alpha$ image shows a similar structure
  with two pairs of faint bipolar lobes oriented at different angles
  \citep[see image displayed by][]{kwoksu03}.  The lobes are not seen
  in the radio continuum because of relatively low sensitivity for the
  radio map.  The extinction map shows a similar structure to that of
  the radio map, where the high-extinction region follows the region
  of high brightness in radio.  However, the two extinction peaks fall
  slightly outside of the radio peaks.  A similar situation has been
  found in the dust extinction map of NGC~7027 \citep{walton88}.


\subsection{M~1-61}

  The radio continuum image shows a shell structure with two
  pronounced emission peaks of roughly the same brightness.  In the
  H$\alpha$ image, the eastern peak is weaker than the western peak,
  and there is at least one pair of faint bipolar lobes in the outer
  region \citep[see image displayed by][]{kwoksu03}.  The extinction
  map shows the optical depth is higher in the region of brighter
  radio emission, with a strong extinction peak near the eastern radio
  peak.


\subsection{M~2-43}

  The radio continuum image shows a shell structure, with the southern
  peak brighter than the northern one.  The H$\alpha$ image shows
  similar structure with slightly extended faint emission outside the
  radio emission region.  However, the northern peak of H$\alpha$
  emission is brighter than the southern one.  As a result, although
  the structure of the extinction map mainly follows that of the radio
  map, the optical depth in the southern part is significant higher
  than the northern part.  The extinction peaks also falls slightly
  outside of the radio peaks.


\subsection{M~3-35}

  The radio continuum image shows a deep central emission minimum and
  two surrounding peaks with concave-like contours at roughly the same
  brightness.  The H$\alpha$ image shows two peaks at approximately
  the same location with two pairs of faint bipolar lobes outside
  \citep[see image displayed by][]{kwoksu03}.  The extinction map has
  similar structure to that of the radio map, with the high-extinction
  region coinciding with the concave-like region.


\subsection{Uncertainties in the Optical Depth}

\begin{figure}
\begin{center}
\includegraphics[width=80mm]{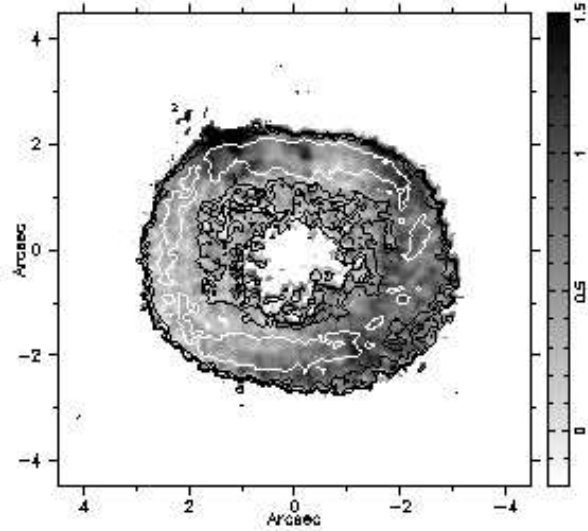}
\caption{Uncertainty optical depth (contours) of (from top to bottom): BD+30$^\circ$3639, IC~5117, M~1-61, M~2-43, M~3-35.
	superposed on the optical depth map in greyscale.  The
	contours levels are BD+30$^\circ$3639: 0.1, 0.2, 0.3, and 0.4, with 0.1 showing
	in white.  IC~5117: 0.02, 0.04,
	0.06, 0.08, 0.1, and 0.2, with 0.02 showing in white.  M~1-61: 0.03, 0.06,
	0.09, 0.12, 0.2, and 0.3, with 0.03 showing in white.  M~2-43: 0.02, 0.05,
	0.1, 0.2, and 0.3, with 0.02 showing in white.  M~3-35: 0.03, 0.06,
	0.09, 0.12, and 0.2, with 0.03 showing in white.}
\label{depth_error}
\end{center}
\end{figure}

  Figure \ref{depth_error}
  show the uncertainties propagated from the radio and optical images.
  The uncertainty associated with the optical depth $\tau$ is given by
\begin{equation}
  \Delta \tau = \sqrt{\left(\frac{\sigma_{\rm radio}}{F_{\rm
  radio}}\right)^2 + \left(\frac{\sigma_{\rm H\alpha}}{F_{\rm
  H\alpha}}\right)^2}
\end{equation}
  where $\sigma_{\rm radio}$ and $\sigma_{\rm H\alpha}$ are the rms
  noise of the radio continuum and optical H$\alpha$ images,
  respectively.  Since the optical images have higher signal-to-noise
  ratios, the uncertainty of the optical depth is dominated by the
  radio image.  Hence the uncertainty map of the optical depth and the
  radio continuum map have similar structure.  The optical depth
  uncertainty ranges from $\Delta \tau \sim$ 0.01 to 0.4, where
  smaller values lie in the regions with stronger radio emission.
  
\placefigure{depth_error}


  The calculations of optical depth above use several assumptions.  To
  investigate whether the optical depth is sensitive to these
  assumptions, the uncertainties propagated from them should also be
  examined.  Among these approximations, the electron temperature
  $T_e$ gives the most significant uncertainty.  While most PNs have
  $T_e$ in the vicinity of 10,000~K, the electron temperatures of PNs
  cover a range of 5,000 - 15,000~K, with $T_e \sim$ 20,000~K in some
  extreme cases.  As a result, different choices of $T_e$ will give
  different values of optical depth.

  To calculate the uncertainty propagated from the assumption that
  $T_e=10,000$~K, we estimate the uncertainty for $T_e$ to be roughly
  $\Delta T_e=\pm 5,000$~K.  
Since
$\tau \propto \ln F_{\rm exp}({\rm H}\alpha) \propto \ln T_e^{-0.53}$,
the uncertainty propagated into the optical depth is
$\Delta \tau \propto 0.53 \times \frac{\Delta T_e}{T_e} \propto 0.53
\times \frac{5,000}{10,000}\sim 0.27$.
This value is an order  of magnitude smaller than the derived optical depth.  
This uncertainty from the electron temperature will only shift the values
of the derived optical depths by a constant, but will not change the
overall appearance of the extinction caused by its dust in a PN.


\subsection{Dust distribution}

  All five PNs in this study show that the H$\alpha$ emission peaks
  and the radio continuum emission peaks are in the same regions.  In
  four of them, the peaks of the dust extinction also reside in the
  same regions.  These results suggest that the ionized gas and the
  dust have similar distributions, so at least part of the dust is
  mixed with the ionized gas.  

On the quantative scale, the faintest parts of the optical image correlate with the peaks of the extinction distribution, as is expected.  
As an independent check, \cite{kastner02} compared the X-ray emission 
  map of BD+30$^\circ$3639 with its extinction map derived from optical and
  infrared images, and showed that the peak of X-ray emission also lies in
  the region with minimum extinction.

We note that the radio images, H$\alpha$ images, and the dust optical depth maps only represent the column density distribution along the line of sight.  In order to remove  projection effects and reconstruct the three
  dimensional ionized gas and dust distributions in PNs,
  we need to develop a 3-D model.
 The details of the model analysis are presented in the next section.

\section{The Three Dimensional Dusty Ellipsoidal Shell (DES) Model}
\label{chap:des}

  While the classification of PN morphology is based on appearance,
  this approach is vulnerable to projection effects.  For example,
  elliptical and bipolar nebulae can appear round when the nebulae are
  viewed pole-on.  In order to study their intrinsic properties, there
  have been many attempts to find the detailed three-dimensional
  structures of PNs \citep{khr68}.  Since many PNs have the appearance of
  limb-brightened elliptical rings that are brightest on their minor
  axes, it is natural to consider models in which the nebulae have an
  elongated, hollow shell structure.  Models with this simple
  geometry, such as the prolate ellipsoidal shell discussed by
  \cite{masson89, masson90}, are successful in reproducing many of the
  shapes observed in PNs.  This was followed by the work of \cite{aaquist96} who developed a
  prolate ellipsoidal shell (PES) model to fit the radio images of PNs.  Their model contains a spherical shell with both radial
  and latitude-dependent density gradients, and the shell is ionized
  by a central star to various depths at different latitudual directions.  The ionized shell, when viewed
  at different angles, results in a variety of morphologies seen in
  PNs.  The parameters derived from the PES model reveal the
  asymmetric distribution of the ionized gas in the nebula.  The PES
  model was later called the ES model and explored extensively by
  \cite{zhang98} to investigate the statistical distributions of the
  asymmetry parameters.  However, because of a typographical error in
  the source code, there are some uncertainties in the derived model parameters in \cite{zhang98}.

  In order to simulate dust extinction effects, we have modified the ES
  model to include a dust distribution with a constant gas-to-dust
  ratio, and we refer to this model as the dusty ellipsoidal shell (DES)
  model.  The DES model has been used to analyze the observational
  images, including radio continuum, H$\alpha$, and dust extinction
  maps to constrain the ionized gas and dust distributions in compact
  PNs.  These different images of the same object provide a way to
  study the distribution of dust in PNs, and also give better
  constraints on the model parameters than a single radio or optical
  image would.  One advantage of using the model is that it helps to
  remove projection effects and to construct the three-dimensional
  structures of PNs.  The DES model will give a quantitative
  classification of PNs morphology rather than the standard
  descriptive terms such as round, elliptical, or bipolar.

\subsection{The Geometry of the DES model}
\label{sec:des}

\begin{figure}
\begin{center}
\resizebox{100mm}{!}{\includegraphics{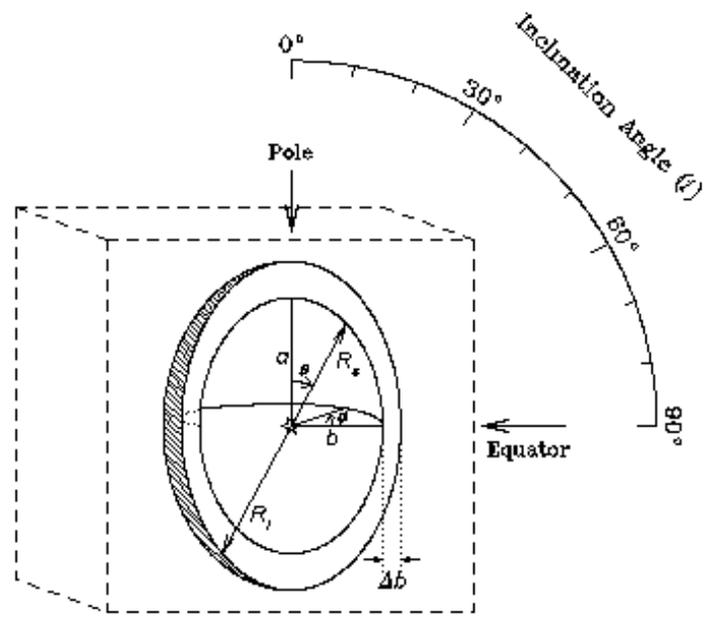}}
\caption[Geometry of DES model] {Geometry of the dusty ellipsoidal
        shell model.  The inclination angle {\it i}, measured in the
        {\it ab} plane, marks the direction toward the observer.}
\label{fig:es}
\end{center}
\end{figure}

  The geometry of the DES model is shown in Figure \ref{fig:es}.  A
  single (hot) star is located inside a shell of ionized gas.  The
  inner region of the nebula is assumed to be empty, supposedly swept
  clean by a wind from the central star.  For simplicity, the empty
  region is assumed to be a prolate or oblate spheroid, depending on
  the choice of polar radius {\it a} and equatorial radius {\it b}.
  The shape of the shell is the result of the interacting winds process, where either the slow AGB wind is equatorially enhanced, or the fast central star wind is non-isotropic (e.g. collimated), or both.  The distance from the star to the inner edge of the gas $R_{s}$ can
  be calculated from {\it a}, {\it b} and the polar angle $\theta$.
  The UV photons from the central star ionize the surrounding gas out
  to the distance $R_{i}$, which depends upon $R_{s}$ and the density
  structure of the surrounding nebula.  

\placefigure{fig:es}

  The model assumes the central star is emitting ionizing photons
  isotropically, and the ionized nebula is ionization-bounded along
  any radial directions from the central star.  
The  depth of the ionization front along any radial vector is given  approximately by 

\begin{equation}
\label{equ:integ}
 \frac{dL}{d\Omega}=\alpha_{B}
   \int_{R_s}^{R_i}n^2(r, \theta, \phi)r^{2}dr~,
\end{equation}
  where $dL/d\Omega$, the number of local ionizing photons ($L$)
  per unit solid angle ($\Omega$), is set equal to the total number of
  recombinations within the local ionized region.  In this equation,
  $\alpha_{B}$ is the recombination coefficient excluding captures to
  the ground level (Case B approximation), $n(r,
  \theta, \phi)$ is the density
  structure of the swept-up shell, and $r, ~\theta, ~\phi$ are the
  spherical coordinates.  $R_{i}$ is a function of $\theta$ and $\phi$
  that depends on the density distribution.  For simplicity, a separable density law is assumed, so that
  $n(r, \theta, \phi) = n_{0} \eta_{r} \eta_{\theta} \eta_{\phi}$;
  here $n_{0}$ is the density at the inner edge, where $(r, \theta,
  \phi)=(R_{s}, 0, 0)$, and $\eta_{r},~\eta_{\theta},~\eta_{\phi}$ are
  dimensionless quantities describing the density relative to $n_{0}$:
\begin{eqnarray}
\label{eq:denr}
 \eta_{r}&=&(r/R_{s})^{-\gamma}  \\ \nonumber \\
 \eta_{\theta}&=& f_\theta \times \left\{ 
  \begin{array}{lr}
   (1-\beta)\displaystyle\left(\frac{2\theta}{\pi}\right)^{\alpha}
	+ \beta,
     & \quad\mbox{if $0 \leq \theta \leq \displaystyle\frac{\pi}{2}$}\\
   \\
   (1-\beta)\displaystyle\left(\frac{2\pi-2\theta}{\pi}\right)^{\alpha}
	+ \beta,
     & \quad\mbox{if $\displaystyle\frac{\pi}{2} \leq \theta \leq \pi$}
  \end{array}\right\} \\ \nonumber \\
 \eta_{\phi}&=& f_\phi~.
\end{eqnarray}
  In this formulation, $\gamma$ gives the logarithmic radial density
  gradient, $\beta$ gives the pole-to equator density ratio, $\alpha$
  governs the shape of the angular density function, and $f_\theta,
  f_\phi$ are additional asymmetry factors:
\begin{eqnarray}
 f_{\theta}&=&\delta_{\theta}+(1-\delta_{\theta})exp[\cos(\theta-\theta_0)-1];
        \ \ \ 0\leq \delta_{\theta} \leq 1 \\
\label{eq:asymphi}
 f_{\phi}&=&\delta_{\phi}+(1-\delta_{\phi})exp[\cos(\phi-\phi_0)-1];  \ \ \
        0\leq \delta_{\phi} \leq 1 ~,
\end{eqnarray}
  where these control asymmetry in the latitude ($\theta$) and
  longitude ($\phi$) directions, respectively.  These two asymmetry
  factors are used to produce the departure from axisymmetry seen in
  many PNs.  For
  simplicity, the phases $\theta_0$ and $\phi_0$ are both assumed to
  be $0^\circ$, so that $f_{\theta}$, $f_{\phi}$ are maximized (=1)
  for $\phi=0^{\circ}$ and $\phi=0^{\circ}$, and minimized for
  $\phi=180^{\circ}$ and $\phi=180^{\circ}$.

  With the given radial density law $\eta_{r}$, eq.~\ref{equ:integ} can be integrated to give a simple analytical
  solution:
\begin{eqnarray}
\label{equ:dL}
 \frac{dL}{d\Omega}&=&\alpha_B \int_{R_s}^{R_i}
  (n_0\eta_r\eta_{\theta}\eta_{\phi})^2 r^2 dr \nonumber \\
  &=&(n_0\eta_{\theta}\eta_{\phi})^2\alpha_B
  \int_{R_s}^{R_i}r^{-2\gamma+2} R_s^{2\gamma}dr \nonumber \\
  &=&(n_{0}\eta_{\theta}\eta_{\phi})^{2}\alpha_{B}
  R_{s}^{3}\frac{(R_{i}/R_{s})^{-2\gamma+3}-1}{-2\gamma+3}~.
\end{eqnarray}

  Once the parameters describing the inner boundary, the density law
  ($\gamma, \eta_{\theta}, \eta_{\phi}$), and the equatorial axis
  shell thickness $\Delta b=[R_{i}-R_{s}]_{\theta=90^\circ}$ are
  chosen, the outer ionized boundary $R_{i}(\theta, \phi)$ can be
  calculated from eq.~\ref{equ:dL}.

  The ionized shell is then projected onto the plane of the sky at
  different inclination angles {\it i}.  For $i=0^\circ$, the nebula
  is seen pole-on, i.e., the line of sight is perpendicular to the
  equatorial plane.  The edge-on case has $i=90^\circ$, with the line
  of sight in the equatorial plane.  The projected emission from the
  ionized region is calculated by assuming that the intensity of the
  recombination H$\alpha$ line and the radio continuum emission is
  proportional to the integration of the square of the density
  function along the line of sight (eqs.~\ref{eq:denr} --
  \ref{eq:asymphi}).  The calculations were done inside a cubic box
  containing the nebula.

  To simulate optical extinction effect, we assume that the dust is well
  mixed with the gas in the ionized region.  Thus, a constant
  gas-to-dust ratio is assigned throughout the ionized region, {\it
  i.e.}, the amount of extinction is directly proportional to the
  density function at each integration point.  The H$\alpha$ emission
  is then attenuated by the dust extinction using eq.~\ref{eq:absorption}.

\subsection{The Model Images}
\label{sec:model}

\begin{figure}
\begin{center}
\plotone{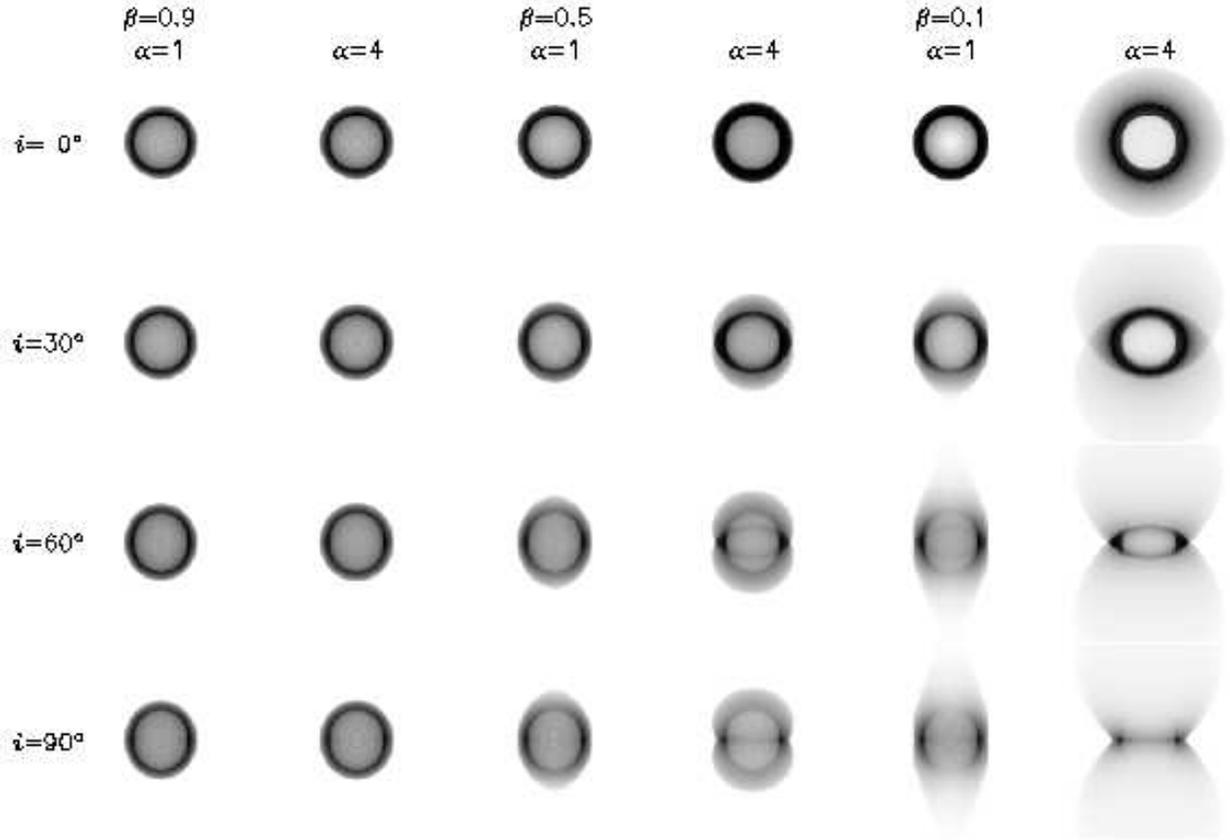}
\caption[Model radio images] {Model radio images assuming no radial
	density gradient ($\gamma=0$).  The images from top to bottom
	show the effect of inclination for $i=0^{\circ}, 30^{\circ},
	60^{\circ}$ and $90^{\circ}$ with different combinations of
	$\beta$ and $\alpha$.  The first pair of columns has
	$\beta=0.9$, with $\alpha=1$ and 4, respectively.  The second
	and third pairs have $\beta=0.5$ and 0.1, also with $\alpha=1$
	and 4.  All images are on a linear grey scale.}
\label{fig:radio}
\end{center}
\end{figure}

\begin{figure}
\begin{center}
\plotone{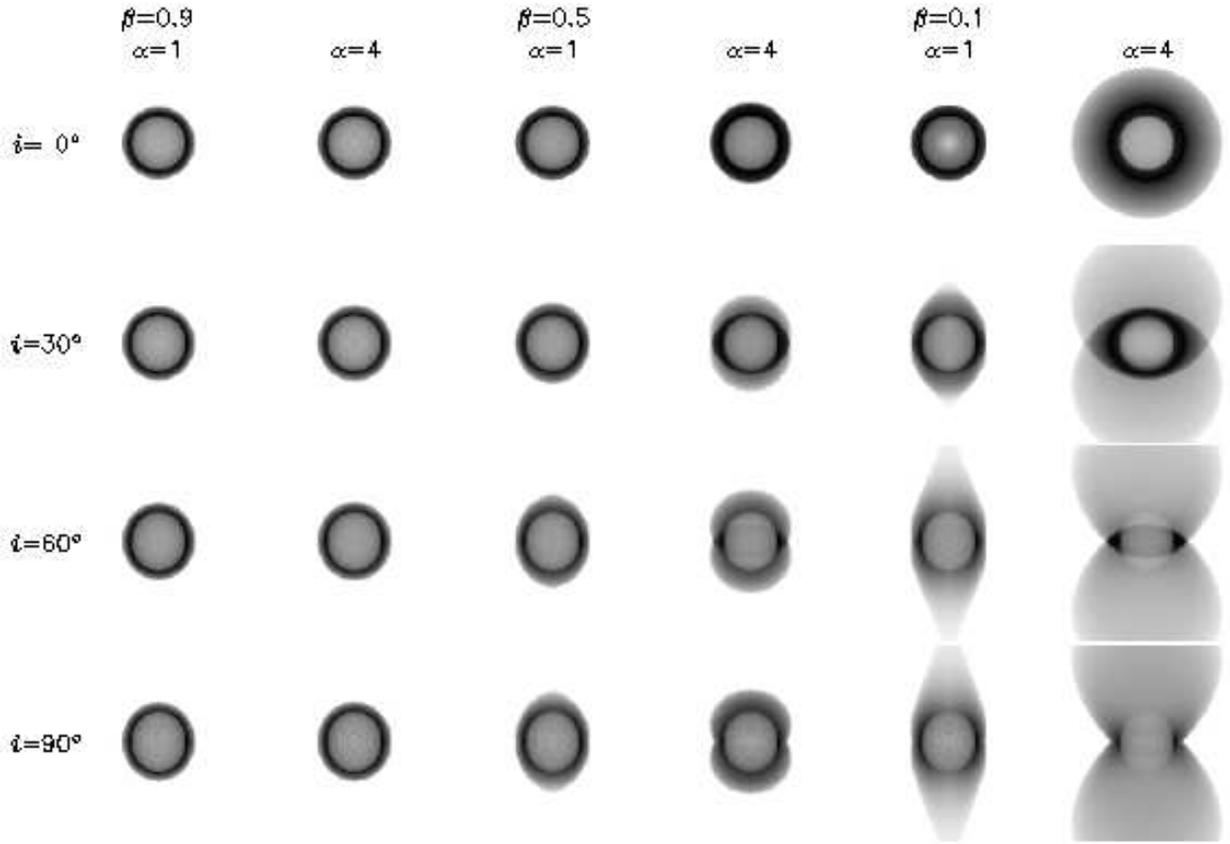}
\caption[Model dust images] {Input model dust distributions using the
        same parameters as in Figure \ref{fig:radio}.  Each dust
        distribution simulates the dust extinction at optical
        wavelengths, although only the dust in front of the emitting
        gas will contribute to the optical extinction (see text for
        discussion).}
\label{fig:dust}
\end{center}
\end{figure}

\begin{figure}
\begin{center}
\plotone{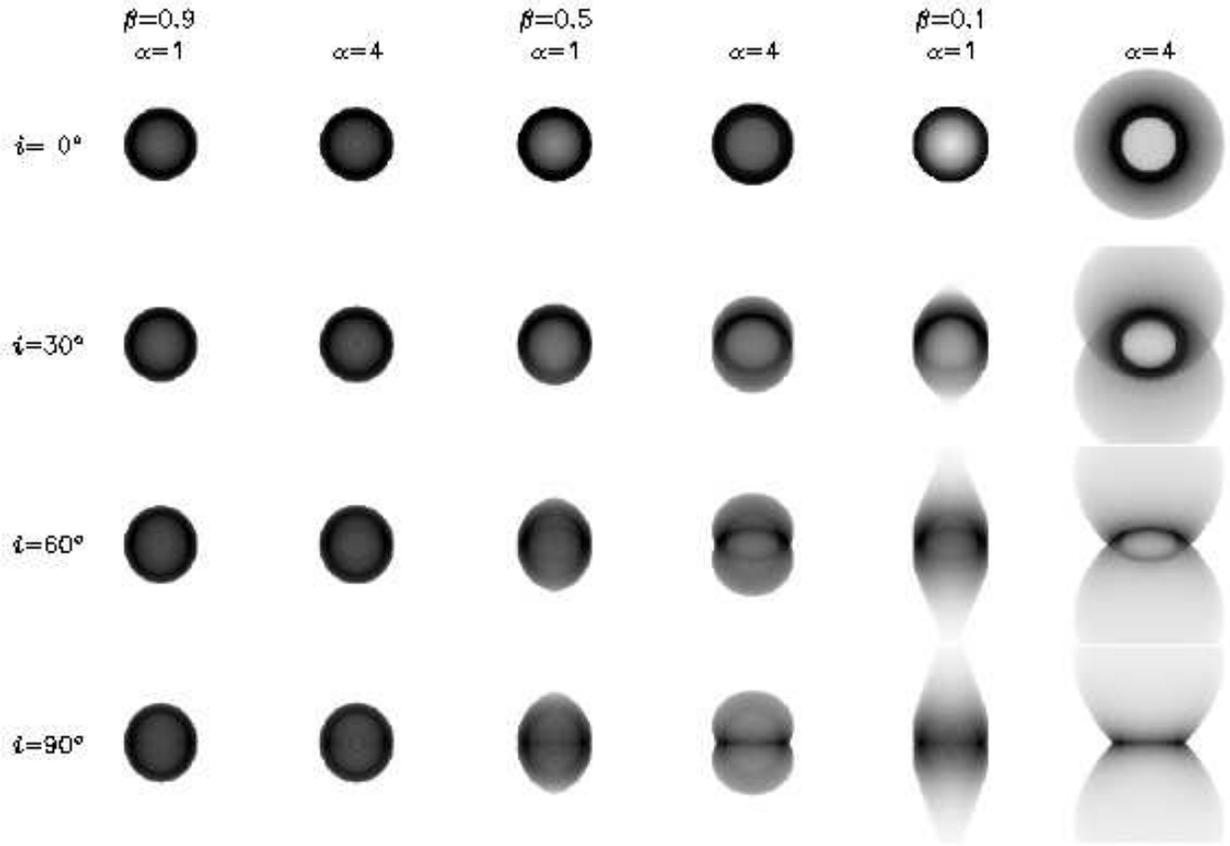}
\caption[Model optical images] {Model optical images with dust
        extinction included.  The parameters are the same as in Figure
        \ref{fig:radio}.}
\label{fig:optic}
\end{center}
\end{figure}

\begin{figure}
\begin{center}
\plotone{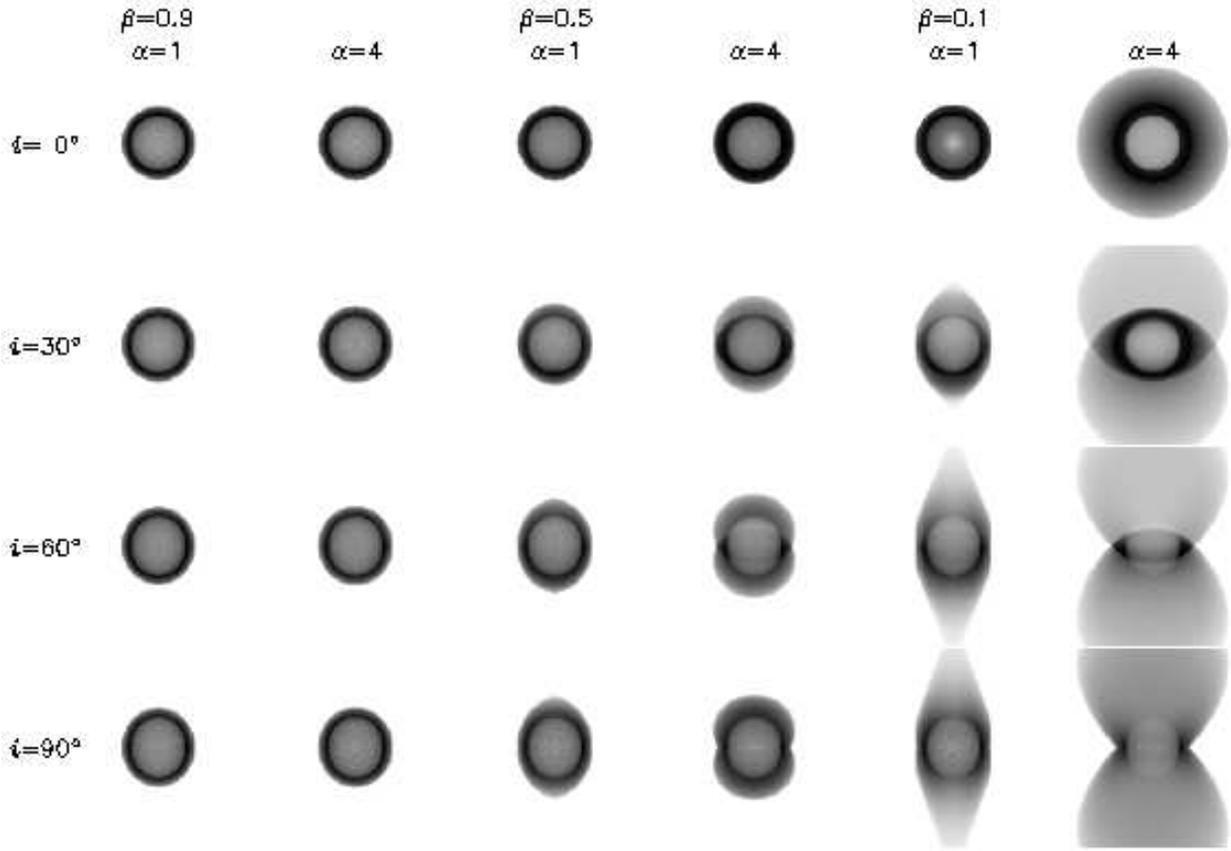}
\caption[Derived extinction images] {Derived extinction images. The
        optical depths are calculated by taking the natural logarithm
        of the ratio of the model radio and optical images. The images
        are arranged in the same order as in Figure \ref{fig:radio}.}
\label{fig:extin}
\end{center}
\end{figure}

  The DES model allows us to explore how the morphology of a nebula
  changes with various parameters, especially $\alpha$ and $\beta$.  A
  series of $128 \times 128$ pixel model images has been produced with
  fixed parameters of $a=20, b=18$, and $\Delta b=6$.  The model
  images presented here have no radial density gradient, {\it i.e.},
  $\gamma=0$.  This is a fairly good approximation, because most
  swept-up shells are thought to be relatively thin, allowing little
  density change with radius.  We have computed the projected images
  for inclination angles $i=0^{\circ}, 30^{\circ}, 60^{\circ}$, and
  $90^{\circ}$ with different combinations of $\beta$ and $\alpha$ as
  shown in Figures \ref{fig:radio} -- \ref{fig:extin}.

  Figure \ref{fig:radio} shows the model images with emission
  calculated from the integration of the square of the density
  function along the line of sight.  This simulates the radio
  continuum images of the long-wavelength free-free emission not
  affected by dust.

\placefigure{fig:radio}

  Figure \ref{fig:dust} displays the input dust distribution. The
  images are computed from the integration of the amount of extinction
  along the line of sight all the way through the nebula.  Because the
  gas-to-dust ratio is constant inside the ionized region, the dust
  distributions appear similar to the model radio images in Figure
  \ref{fig:radio}.  
If the dust  temperature is uniform, and the cloud is optically thin for the dust
  thermal emission, the flux emitted by the dust is directly proportional
  to the absorption coefficient.  Therefore, these dust distributions
  can mimic infrared images of the dust emission.

\placefigure{fig:dust}

  Figure \ref{fig:optic} gives the model images when the dust
  extinction is taken into account.  As in Figure \ref{fig:radio},
  emission is calculated from the integration of the square of the
  electron density, but now it is attenuated by the dust in front of
  it along the line of sight.  This simulates the optical H$\alpha$
  images, which, unlike the radio continuum maps, are affected by
  dust.  Some images clearly show an asymmetric appearance caused by
  dust extinction (e.g., the image of
  $\alpha=1,~\beta=0.1,$~and~$i=30^\circ$).

\placefigure{fig:optic}

  Figure \ref{fig:extin} displays output extinction optical depth
  maps, calculated as the natural logarithm of the ratio of the model
  radio and optical images (eq.~\ref{eq:tau}).  Because only dust in
  front of the emission can attenuate the light, the amount of dust in
  the extinction maps (Fig.~\ref{fig:extin}) is less than the actual
  amount of dust in the nebulae (Fig.~\ref{fig:dust}).  Comparing these
  images with the input dust distributions in Figure \ref{fig:dust},
  the amount of missing dust for different inclination angles and
  morphologies can be estimated.  By comparing the radio and H$\alpha$
  images, the nebular geometry can be estimated and the asymmetry
  effect caused by dust for this viewing geometry can be removed and
  reveal any intrinsic asymmetry of the gas distribution.

\placefigure{fig:extin}

\section{Comparison of Observed and Model Images}
\label{sec:compare-des}

\begin{figure}
\begin{center}
\includegraphics[width=\textwidth]{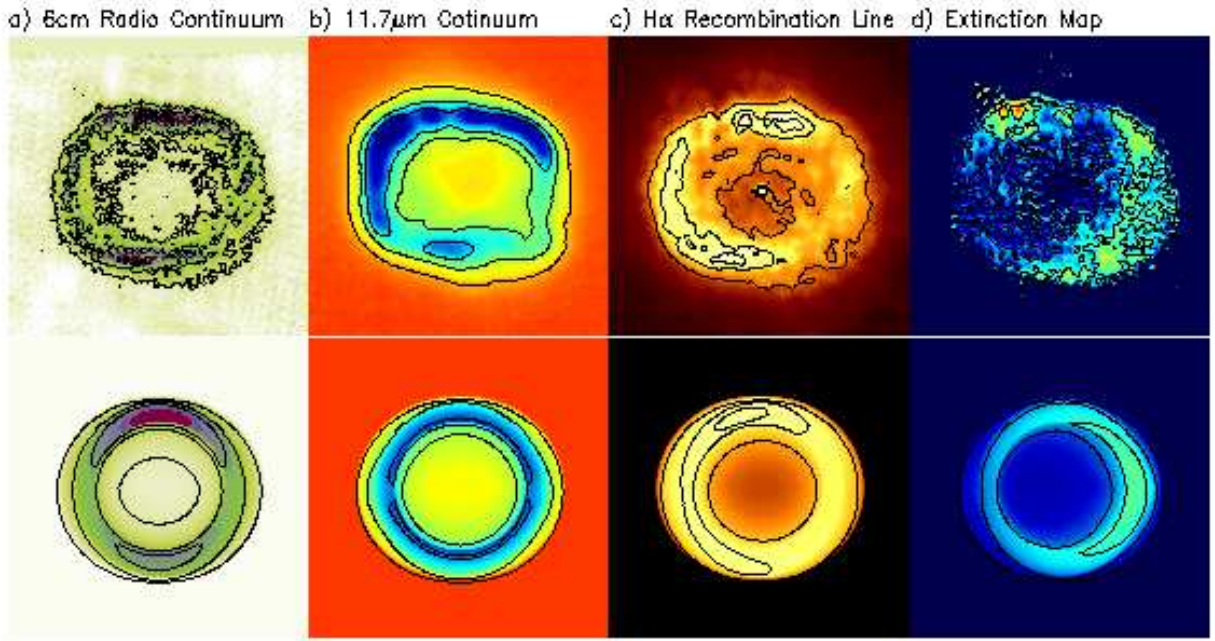}
\caption{Comparison of the model with the observed
	images for BD+30$^\circ$3639.  The observed images are in the
	top panels and the model images are in the bottom panels.  The
	images are: a) 6-cm radio continuum emission \citep{bryce97}.
	b) Gemini 11.7$\mu$m emission \citep{volk03}.  c) {\it HST}
	H$\alpha$ emission (from {\it HST} archive).  d) Extinction
	map derived from the radio continuum and H$\alpha$ emission.   All displays are on a linear scale except the {\it HST} observed and model images are in square-root of intensity.}
\label{fg:bd+30.model}
\end{center}
\end{figure}

\begin{figure}
\begin{center}
\includegraphics[width=\textwidth]{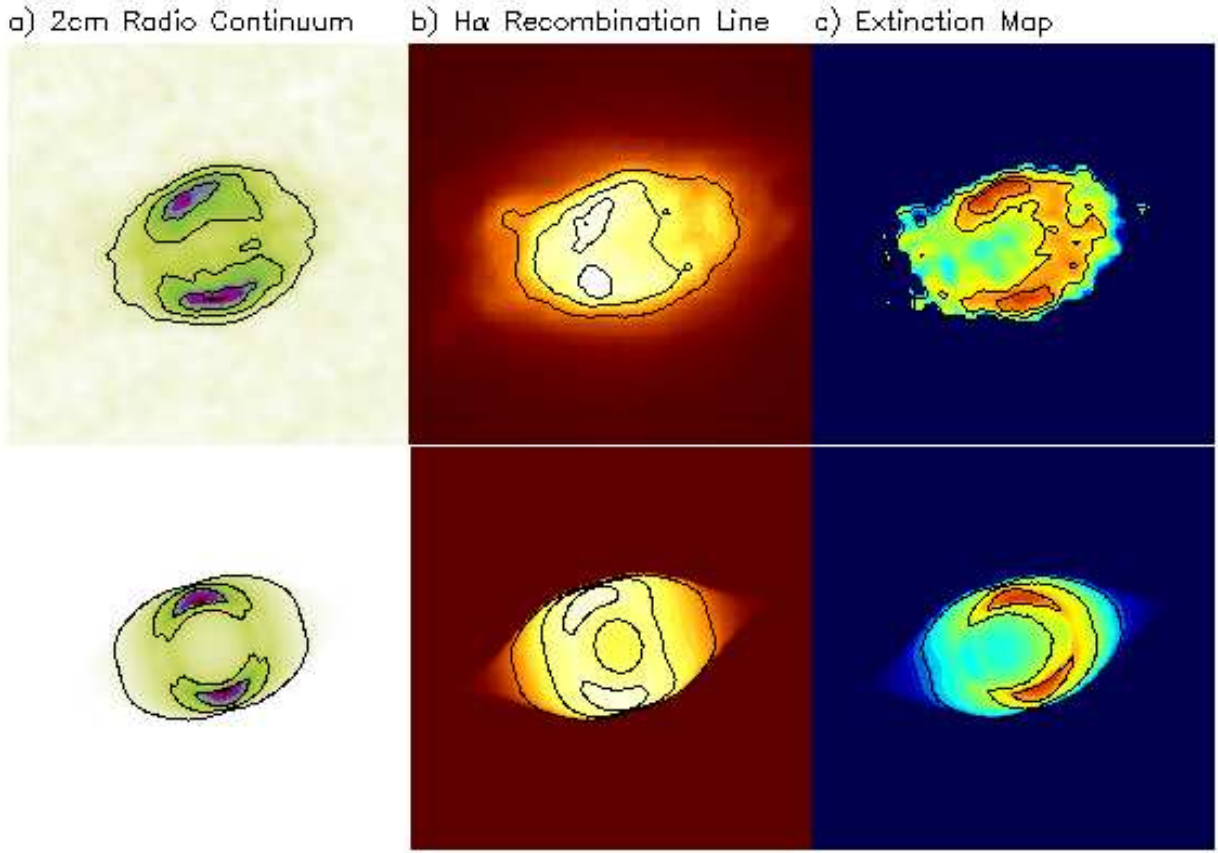}
\caption[Comparison of the model with the observed images for IC~5117]
	{Comparison of the model with the observed images for IC~5117.
	The observed images are in the top panels and the model images
	are in the bottom panels.  The images are: a) $\lambda$ 2 cm radio
	continuum emission.  b) {\it HST} H$\alpha$ emission
	\citep{kwoksu03}.  c) Extinction map derived from the radio
	continuum and H$\alpha$ emission.  The radio and extinction images are displayed on a linear scale whereas the {\it HST} images are in square-root scale.}
\label{fg:ic5117.model}
\end{center}
\end{figure}

\begin{figure}
\begin{center}
\includegraphics[width=\textwidth]{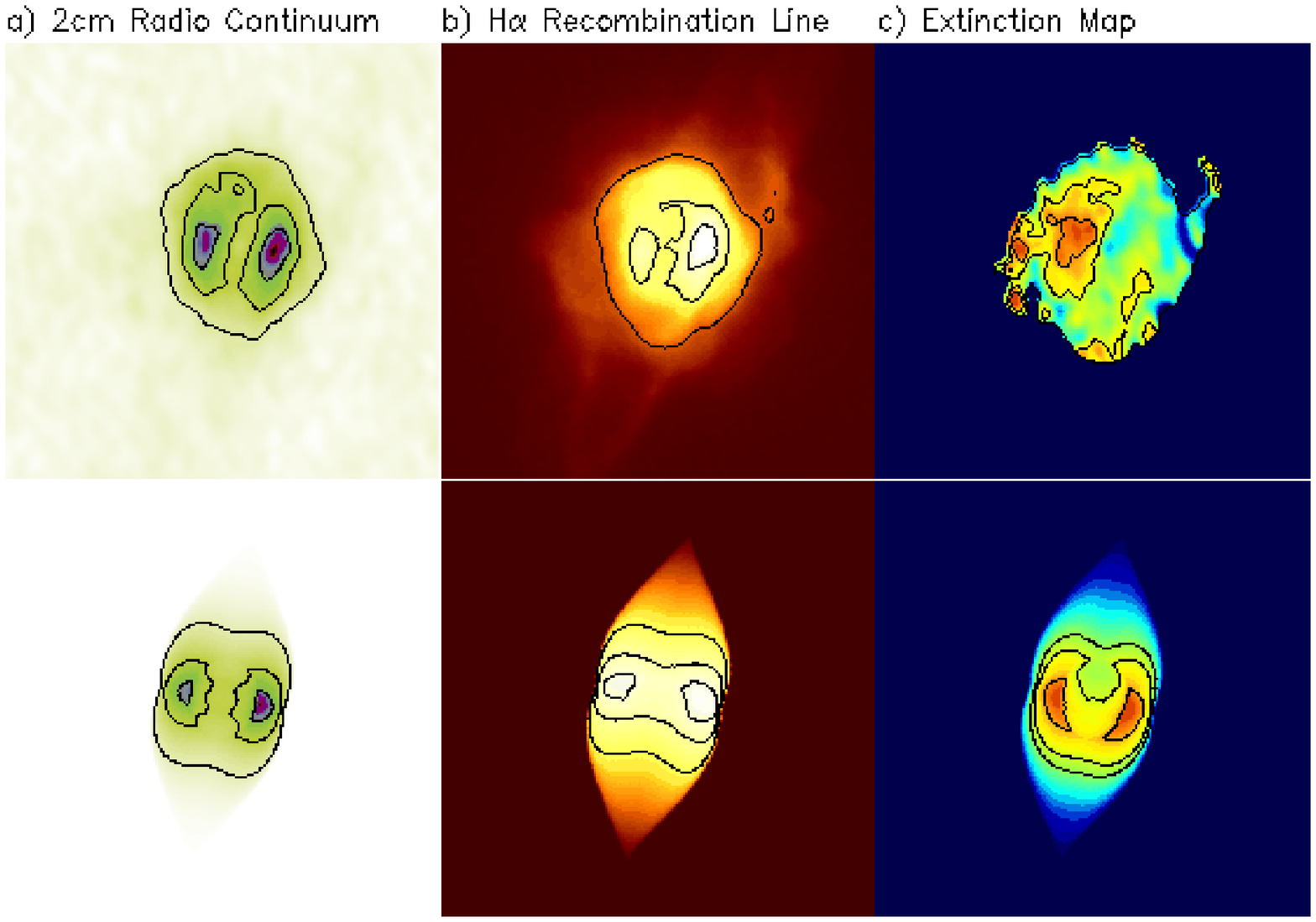}
\caption[Comparison of the model with the observed images for M~1-61]
	{Comparison of the model with the observed images for M~1-61.
	The images are arranged in the same manner as
	Fig. \ref{fg:ic5117.model}.}
\label{fg:m1-61.model}
\end{center}
\end{figure}

\begin{figure}
\begin{center}
\includegraphics[width=\textwidth]{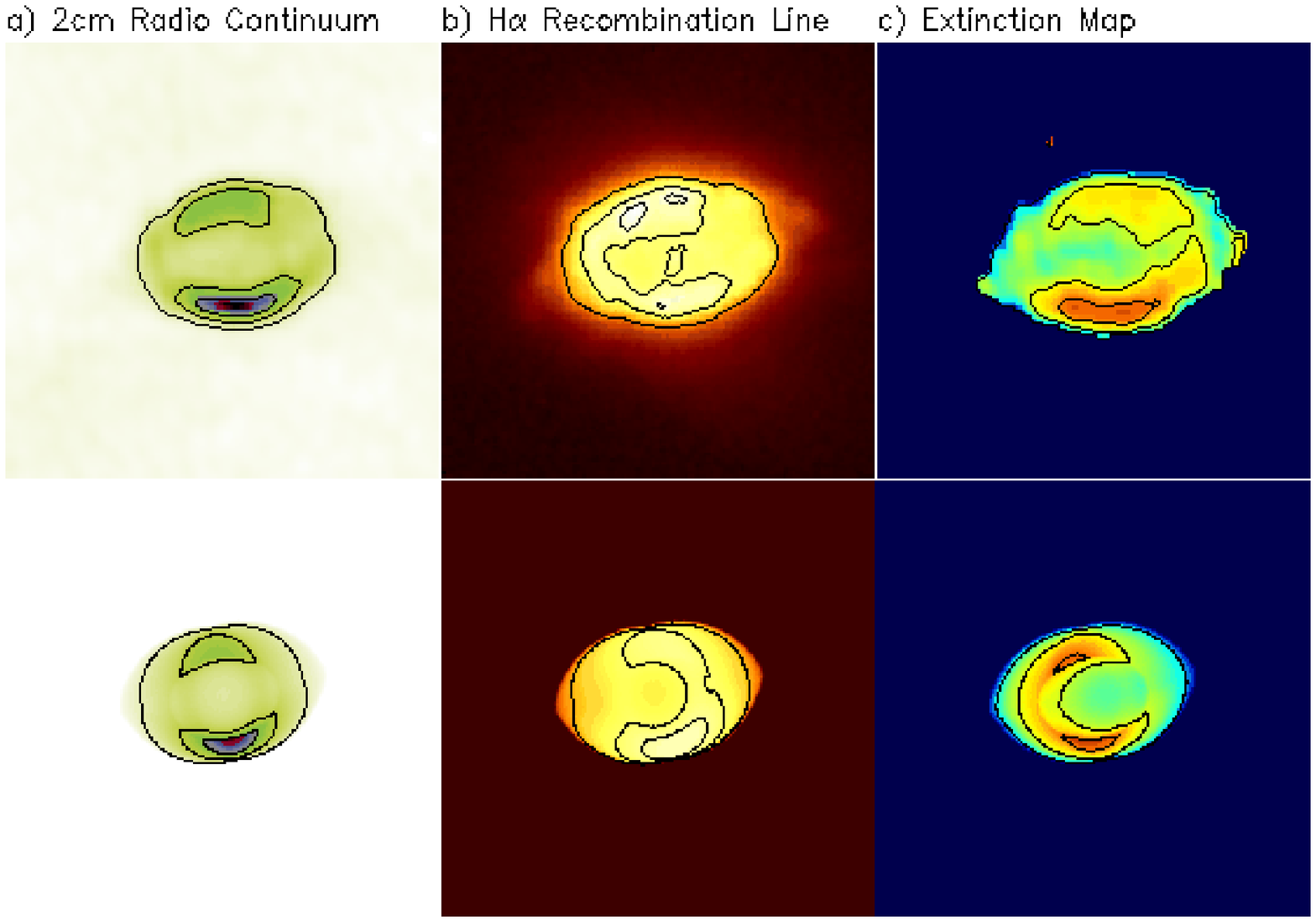}
\caption[Comparison of the model with the observed images for M~2-43]
	{Comparison of the model with the observed images for M~2-43.
	The images are arranged in the same manner as
	Fig. \ref{fg:ic5117.model}.}
\label{fg:m2-43.model}
\end{center}
\end{figure}

\begin{figure}
\begin{center}
\includegraphics[width=\textwidth]{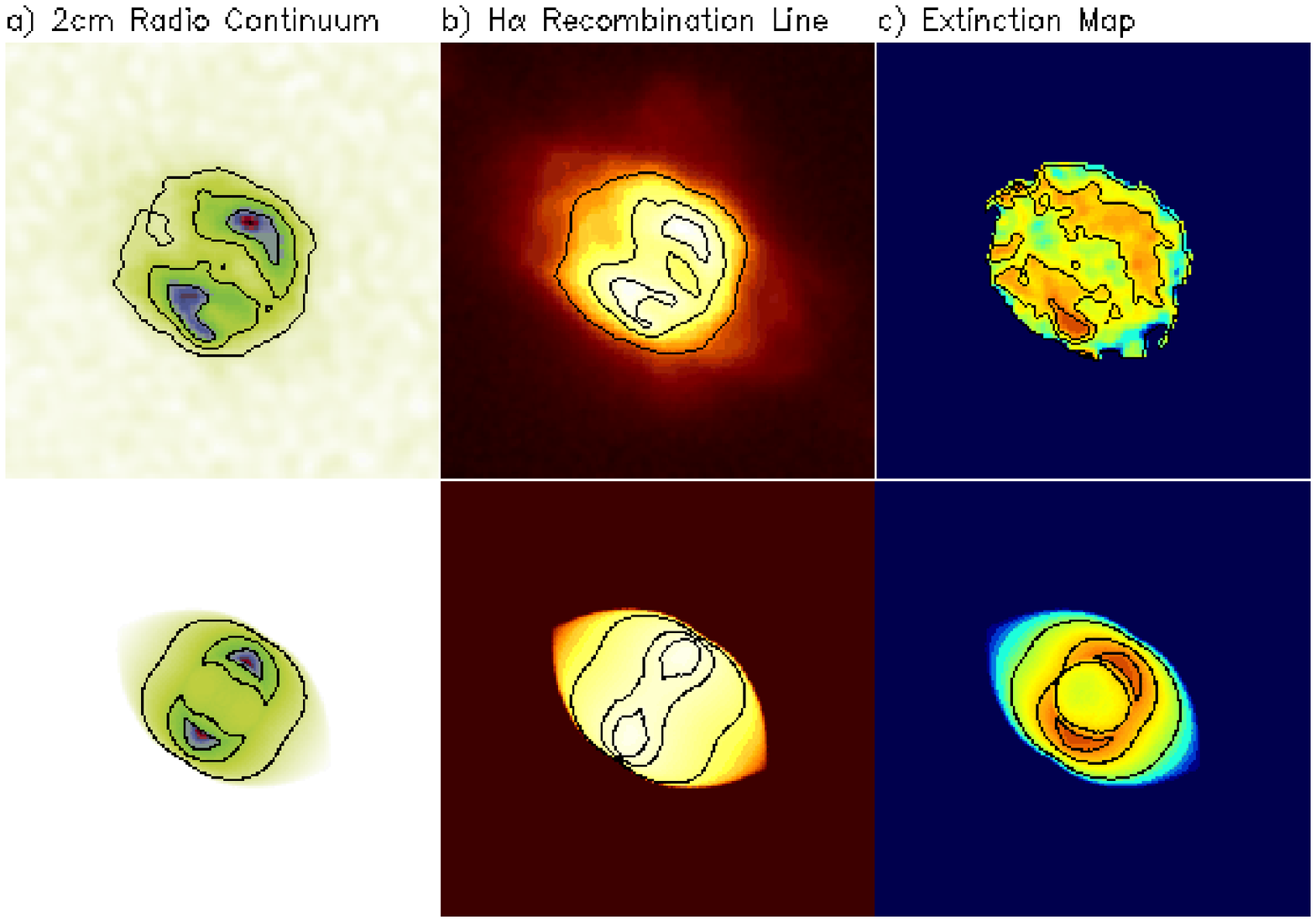}
\caption[Comparison of the model with the observed images for M~3-35]
	{Comparison of the model with the observed images for M~3-35.
	The images are arranged in the same manner as
	Fig. \ref{fg:ic5117.model}.}
\label{fg:m3-35.model}
\end{center}
\end{figure}

  In order to constrain the parameters that describe the geometry and
  angular density distribution of the PNs presented, a
  set of model images were made to compare to each individual nebula's
  observed radio continuum and H$\alpha$ images, extinction maps, and
  in one case (BD+30$^{\circ}$3639), an infrared dust emission image
  \citep[from][]{volk03}.  The comparison was made by visual
  inspection.  The faint emission from the outer parts of each nebula
  was chosen to match the observed H$\alpha$ images because their S/N
  ratios are much higher than those for the radio continuum images.
  For each object, a set of parameters was found that generate model
  images matching most of the observed images.  The images comparing
  the best models to the observations for these five nebulae are shown
  in Figures \ref{fg:bd+30.model} -- \ref{fg:m3-35.model}.  The
  parameters used to generate the model images are listed in Table
  \ref{tb:espar}.

\subsection{BD+30$^\circ$3639}

  The model of BD+30$^\circ$3639 has a relatively small equatorial
  thickness $\Delta b$ compared to the size of the inner empty region,
  which has a prolate shape with an axial ratio of $a/b=1.2$.  The small value of $\beta(=0.1$) suggests that it has a large equatorial to polar density ratio.  A small asymmetry
  $\delta_{\phi}=0.9$ in the longitude direction is added to explain
  the unequal intensity of the two sides observed in the radio image.
  Dust well-mixed with ionized gas is able to reproduce the
  anti-correlation of the optical image and extinction map.  
Due to its low inclination angle  ($i=15^\circ$), it has a round appearance.  This
  low inclination is consistent with the value of $i\sim10^\circ$
  found by \cite{li02}, who also fit the radio image and the spectra
  by approximating the nebula as an ellipsoid.  But a comparison with the model image in Fig.~\ref{fig:optic} (second column from the right) suggests that BD+30$^\circ$3639 would appear as a prominent bipolar nebula if viewed edge-on.  

\placetable{tb:espar}

\placefigure{fg:bd+30.model}

\subsection{IC~5117}

  The model of IC~5117 also has a relatively small equatorial
  thickness compared to the inner empty region, with a prolate axial
  ratio $a/b=1.67$.  It has a small pole-to-equator ratio of
  $\beta=0.1$.  No asymmetry factor in the longitude direction is
  needed.  The brighter region toward the east in the optical image
  and the higher extinction toward the west can be reproduced
  by a dust distribution with a constant gas-to-dust ratio.  The model
  images have an intermediate inclination angle of $i=40^\circ$.

\placefigure{fg:ic5117.model}

\subsection{M~1-61}

  The model of M~1-61 has an equatorial thickness comparable to the
  dimension of the inner empty region, which has a prolate axial ratio
  of $a/b=1.2$.  It has a small pole-to-equator ratio of $\beta=0.2$.
  A small asymmetry $\delta_{\phi}=0.9$ in the longitude direction is
  needed to explain the mildly brighter western peak in the radio
  image.  The anti-correlation of the optical bright peak in the west
  and the strong extinction peak in the east can be reproduced by the
  well-mixed dust.  However, it is not as prominent as the observed
  images.  The model images has a high inclination angle of
  $i=70^\circ$, and they suggest that a higher dynamic range image of M~1-61 will reveal a pair of bipolar lobes.

\placefigure{fg:m1-61.model}

\subsection{M~2-43}

  The model of M~2-43 also has a relatively small equatorial thickness
  compared to the inner empty region with a prolate axial ratio
  $a/b=1.67$.  It has a relatively small pole-to-equator ratio of
  $\beta=0.3$.  A mild asymmetry $\delta_{\phi}=0.7$ in the longitude
  direction is added to reproduce the brighter southern peak in the
  radio image.  Although a simple constant gas-to-dust ratio is able
  to generate an anti-correlation between the H$\alpha$ emission and
  optical depth, it cannot reproduce optical and extinction maps that
  match the observed images well.  The model images have an
  intermediate inclination angle of $i=45^\circ$.

\placefigure{fg:m2-43.model}

\subsection{M~3-35}

  The model of M~3-35 has an equatorial thickness comparable to the
  dimension of the inner empty region, which has an oblate axial ratio
  of $a/b=0.83$.  This is the only object that requires an oblate
  model.  It has a intermediate pole-to-equator ratio of $\beta=0.4$.
  No asymmetry factor in the longitude direction is needed to
  reproduce the radio image.  No obvious anti-correlation between
  optical and extinction maps can be found.  The model images are
  highly inclined to the line of sight, with $i=80^\circ$.  As in M~1-61, a higher dynamic range image is expected to reveal the bipolar lobes in the NE-SW directions.

\placefigure{fg:m3-35.model}

\section{Discussion}
\label{sec:discussion-des}

  As we can see from Figures \ref{fg:bd+30.model} -- \ref{fg:m3-35.model},
  the model images can generally simulate the observed images.  The remaining differences could be accounted for by several effects.  For example, the assumed
  model geometry may be too simple to reproduce all of the complex
  small-scale structure of PNs.  Other causes could include
  clumpy dust distributions that do not follow a simple constant
  gas-to-dust ratio.   Three out of five PNs show an asymmetric density distribution in the longitudinal direction, a departure from axisymmetry. As summarized by \cite{soker01}, four main processes can result in
  the departure from axisymmetry in PNs: interaction
  with the interstellar medium, mass-loss enhancements due to cool
  starspots, or a wide or a close binary companion.  They investigated the effects of companions and estimated that about 27\% of all PNs are expected to acquire 
non-axisymmetric structure from binary interactions. 
Nevertheless, the effects of the  ISM are also worth considering.  Since the majority of PNs are  distributed near the Galactic plane where the mean density of the
  ISM is higher, significant interaction between PNs and the ISM is
  expected.  Such phenomena can be seen from the disturbed emission on
  the outskirts of some nebulae \citep[e.g.,][]{xilouris96} and from
  some apparently off-centered nuclei (e.g., Sh 2-216, Tweedy et al. 1995; KFR 1 and MeWe 1-4, Rauch et al. 2000).

The electron densities of these five PNs are all
  found to be higher near their equators than near their poles.  The
  small pole-to-equator density ratios ($\beta$) of some nebulae
  indicate they would show a bipolar morphology if seen edge-on.  
The fraction of intrinsic bipolar PNs  therefore are likely to be higher than suggested by classification by apparent morphology.  Since the morphology of PNs has been proposed to relate to their physical properties, \citep [e.g., bipolar nebulae have been suggested to correlate with nebulae with more massive central stars;][]{sta93}, it is therefore important that we have a good understanding of the intrinsic strucures of PNs.  By using models such as DES, we can more reliably estimate the fraction of bipolar nebulae and therefore assess any correlation with physical properties of the PNs.

\section{Conclusions}

We have derived the dust distribution in five compact PNs from their H$\alpha$ and radio images.  From this study, we are able to determine the extent of contribution by dust extinction to the apparent departure from axi-symmetry in these PNs.  By comparing these results with simulations generated by the ellipsoidal shell model, we can infer the intrinsic morphologies of the PNs.  This method gives a more quantitative approach to the analysis of PN morphology than visual inspection of apparent morphology, and is therefore more capable to deducing proper statistics on the distribution of morphological types of PNs.

\acknowledgements 

We thank Kate Y.L. Su, Kevin Volk, Orla Aaquist, Steven Gibson for helpful discussions.  Some figures were made with the software package WIP (Morgan 1995).  We are grateful to R. Gooch for his expansion of the capabilities of the 
Karma visualization software package (Gooch 1996)\footnote{See also http://www.atnf.csiro.au/karma}.
This work was supported in part by grants to SK from the Natural Sciences and Engineering Research Council of Canada and by the National Science Council of Taiwan.  SK acknowledges the award of a Killam Fellowship from the Canada Council for the Arts.

\clearpage

\begin{figure}

\begin{center}
\includegraphics[width=\textwidth]{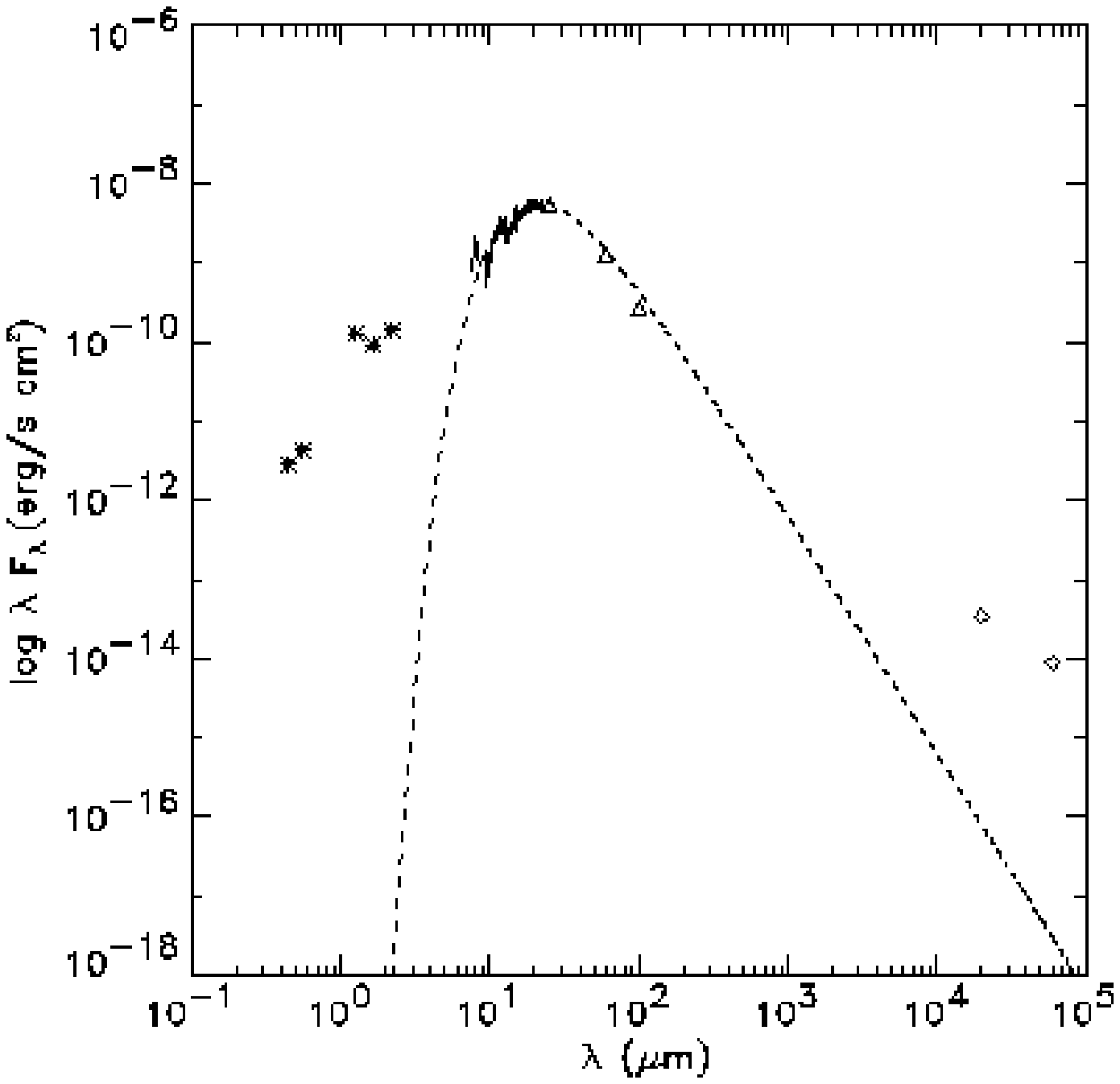}
\end{center}
\end{figure}

\begin{figure}
\begin{center}
\includegraphics[width=\textwidth]{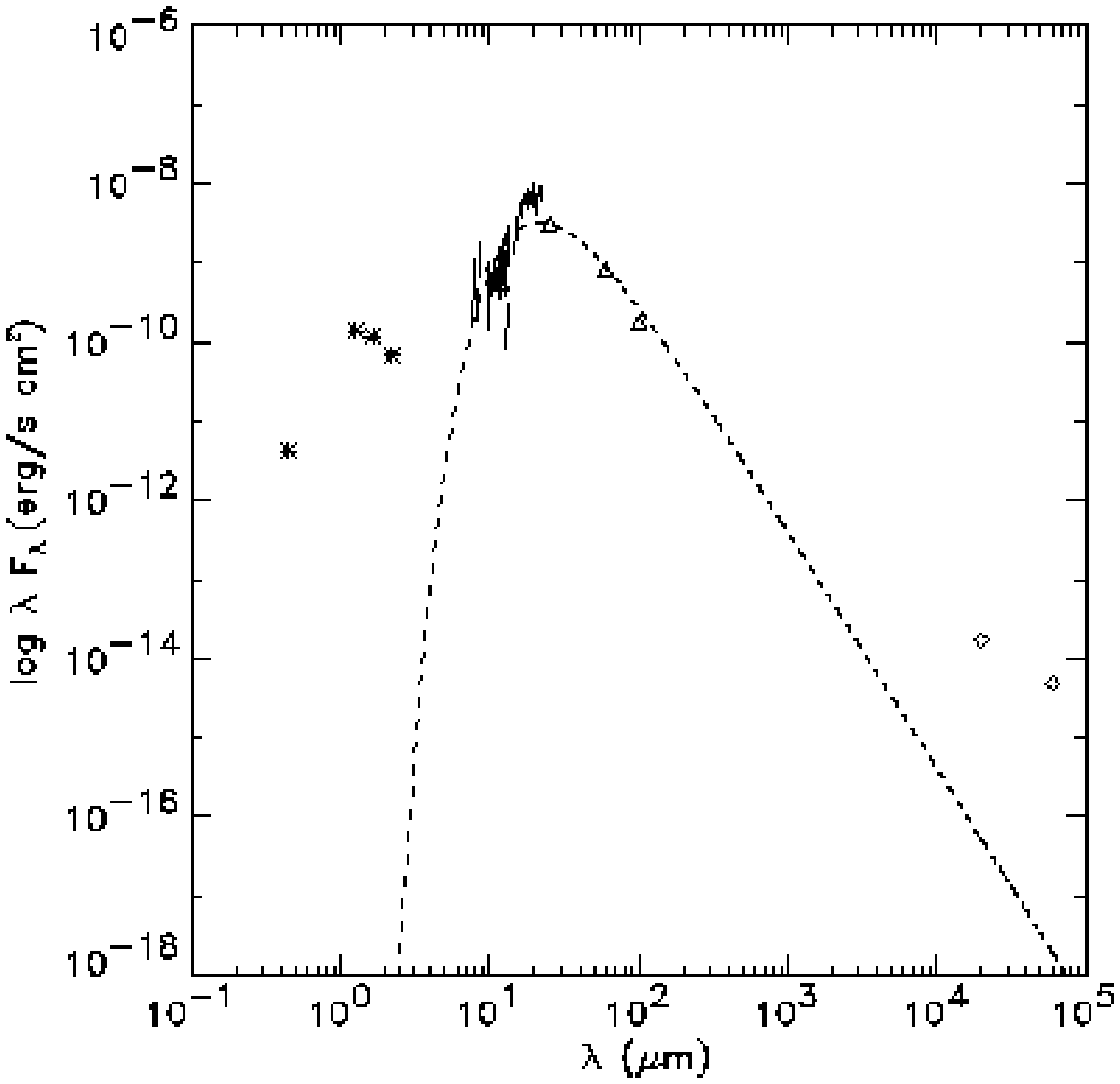}
\end{center}
\end{figure}

\begin{figure}
\begin{center}
\includegraphics[width=\textwidth]{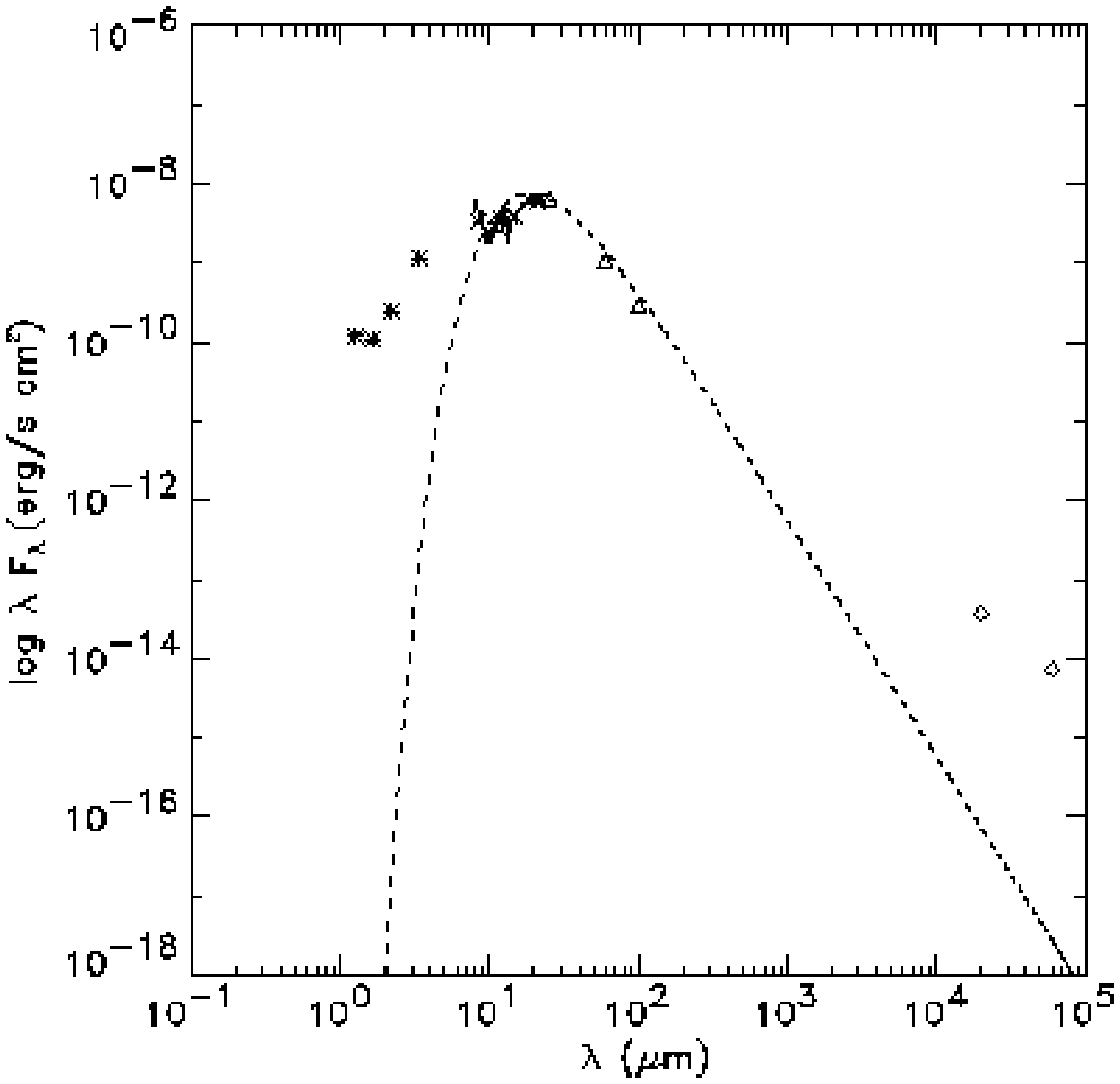}
\end{center}
\end{figure}

\begin{figure}
\begin{center}
\includegraphics[width=\textwidth]{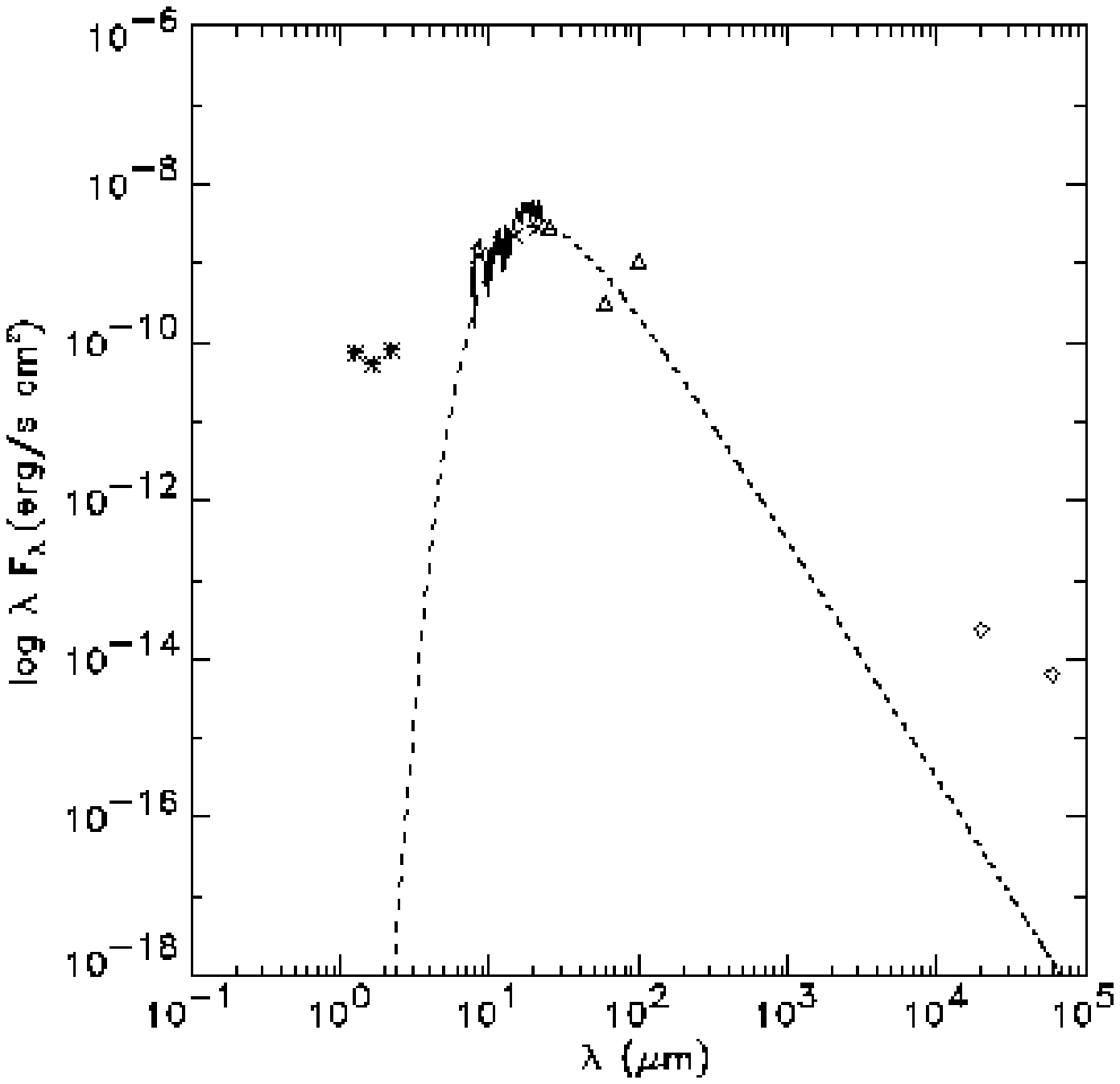}
\end{center}
\end{figure}

\clearpage

\begin{figure}
\begin{center}
\includegraphics[width=\textwidth]{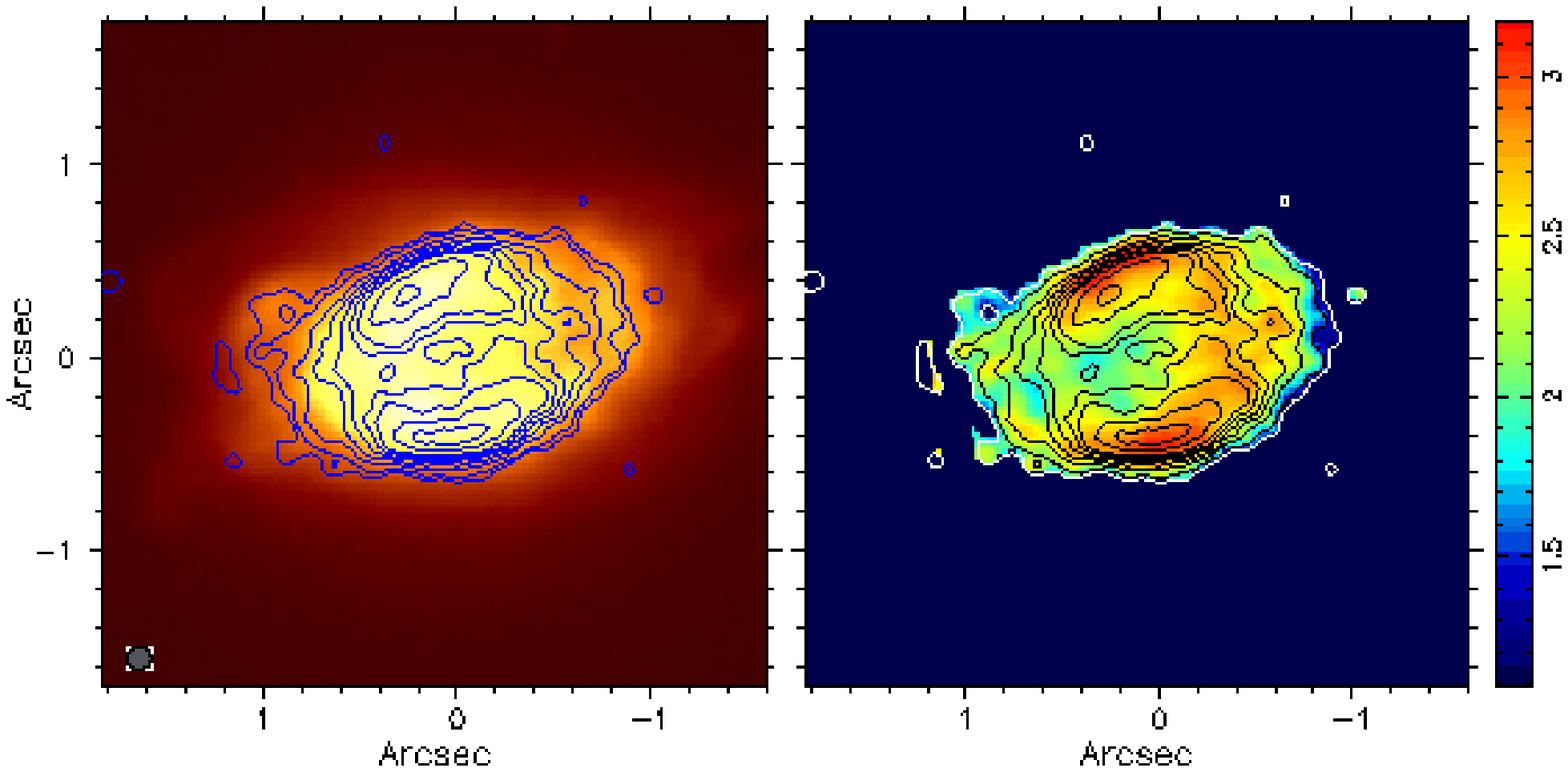}
\end{center}
\end{figure}

\begin{figure}
\begin{center}
\includegraphics[width=\textwidth]{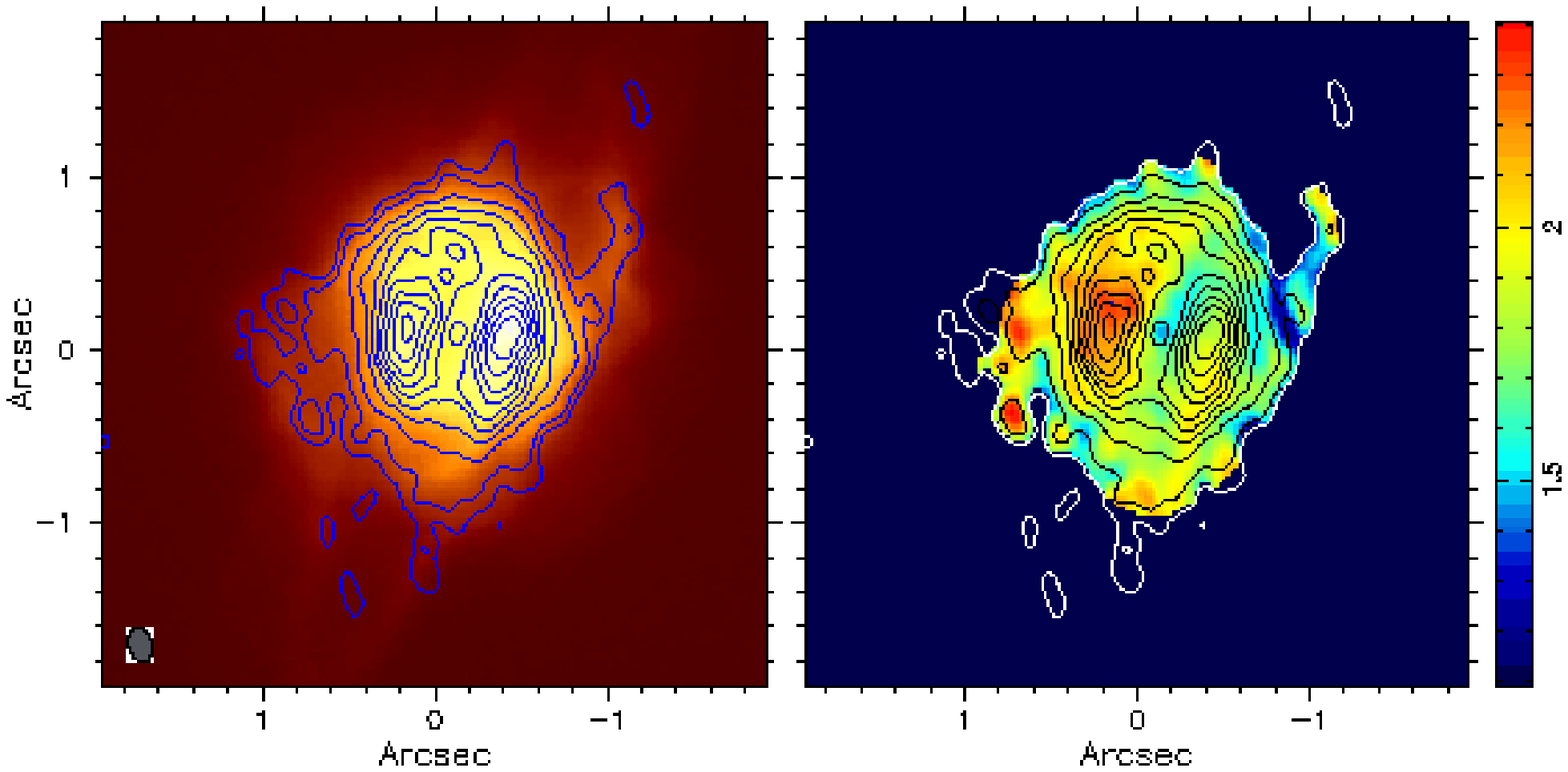}
\end{center}
\end{figure}

\begin{figure}
\begin{center}
\includegraphics[width=\textwidth]{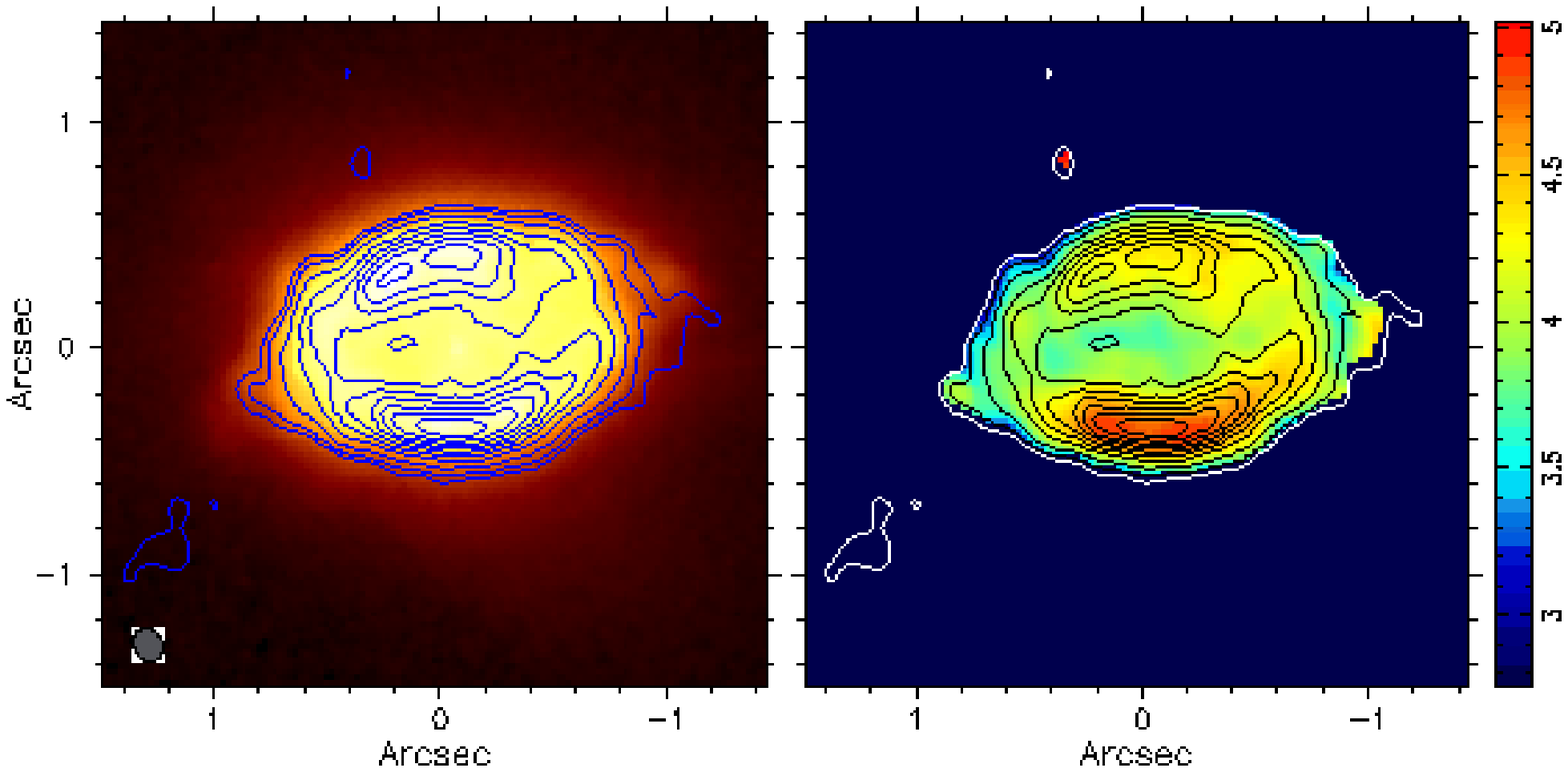}
\end{center}
\end{figure}

\begin{figure}
\begin{center}
\includegraphics[width=\textwidth]{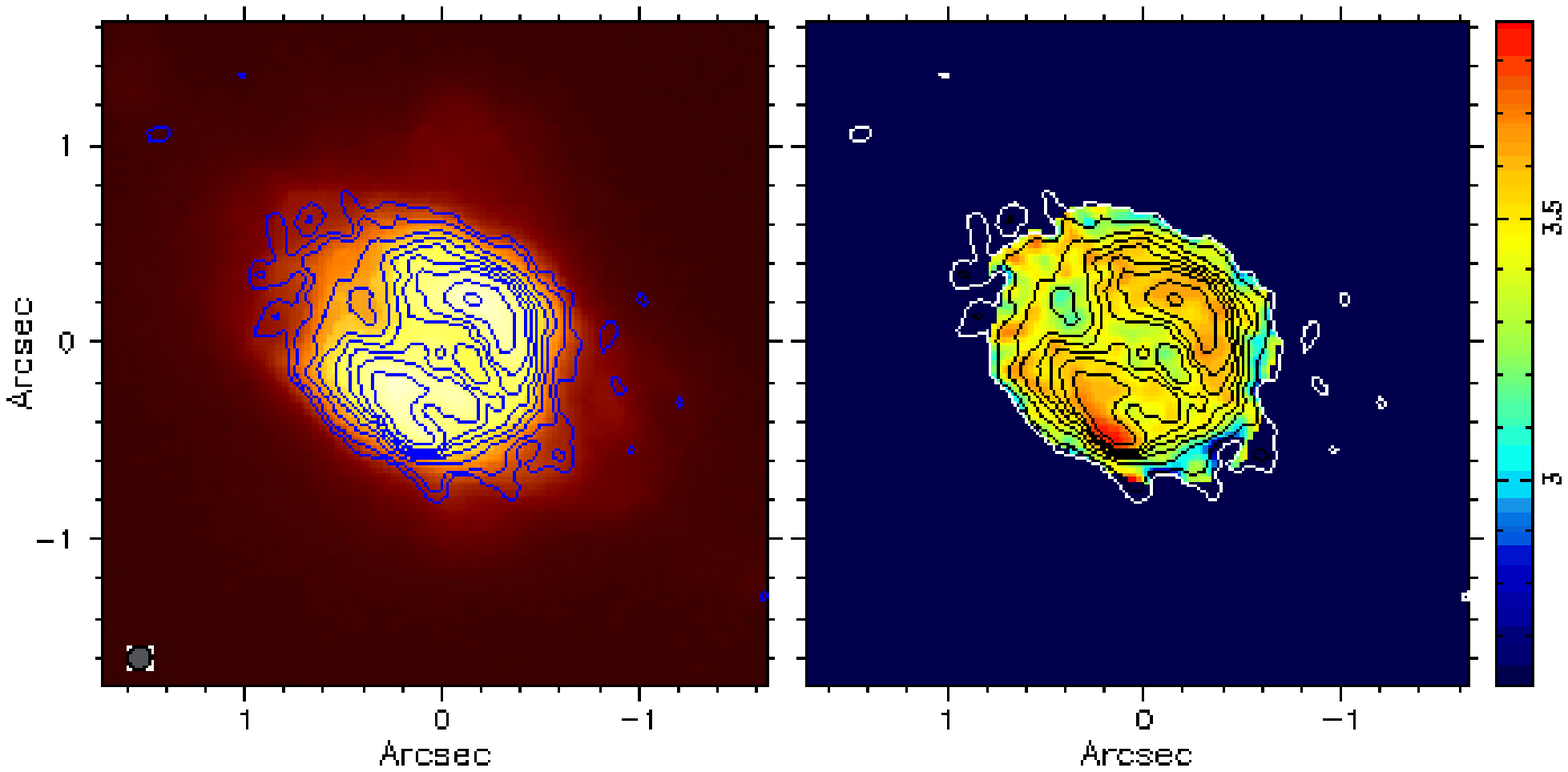}
\end{center}
\end{figure}

\clearpage

\begin{figure}
\begin{center}
\includegraphics[width=80mm]{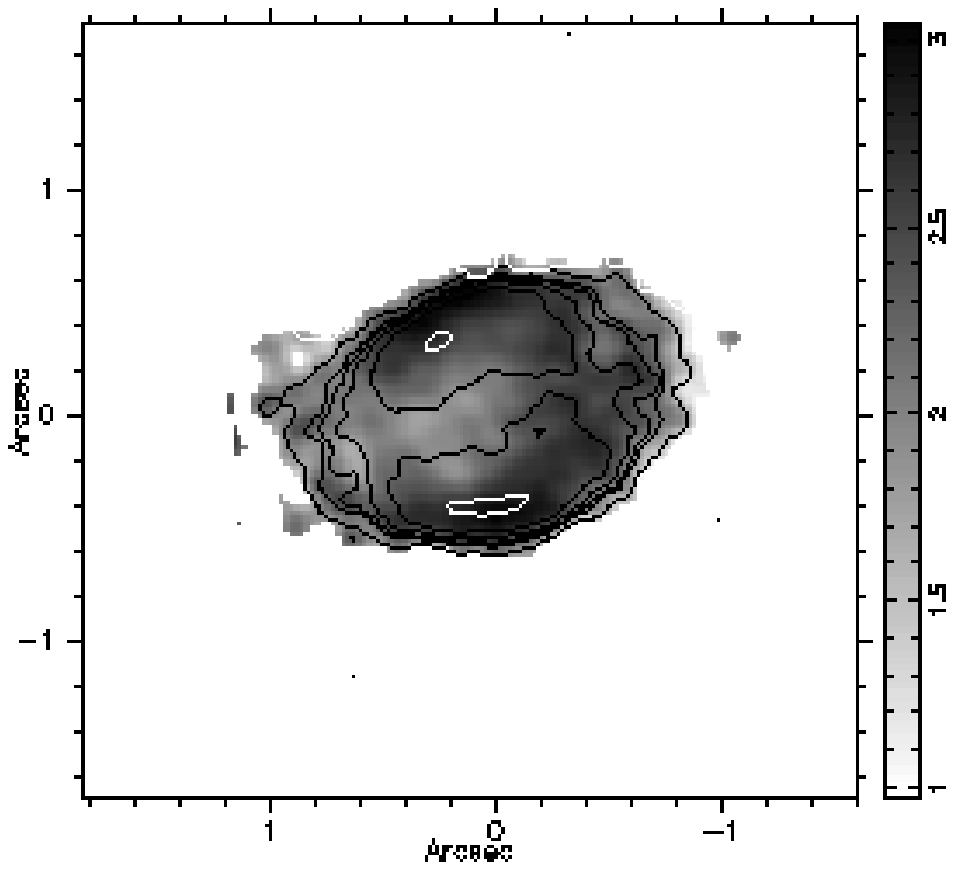}
\end{center}
\end{figure}

\begin{figure}
\begin{center}
\includegraphics[width=80mm]{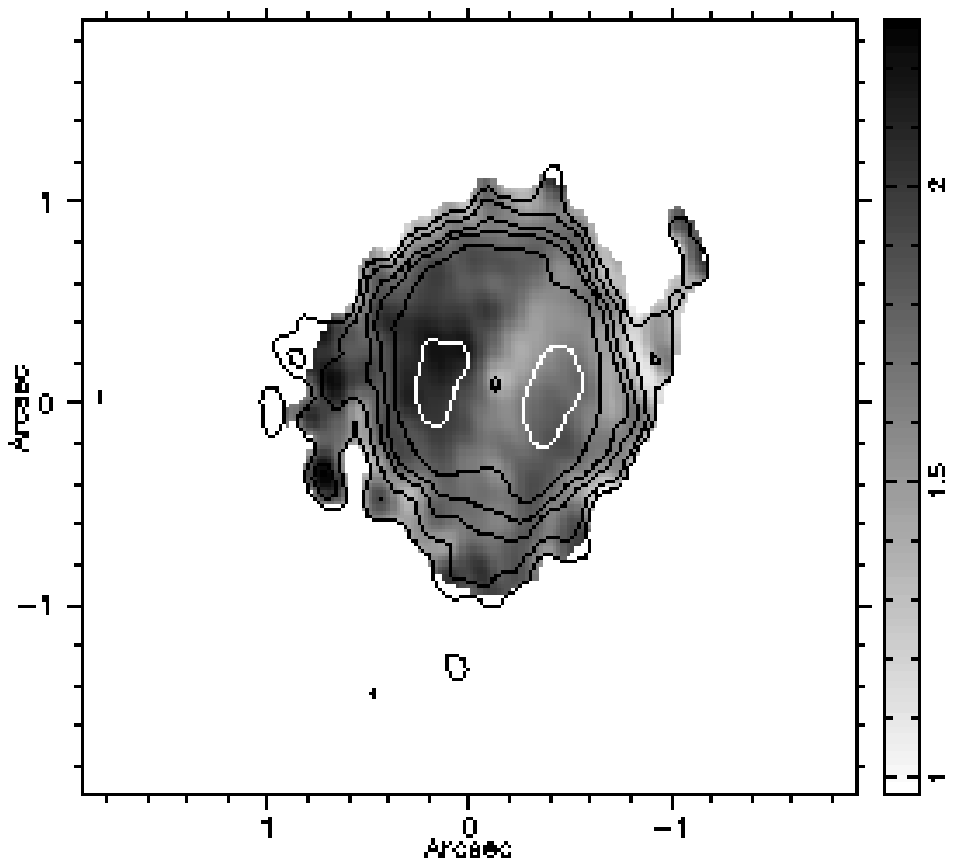}
\end{center}
\end{figure}

\begin{figure}
\begin{center}
\includegraphics[width=80mm]{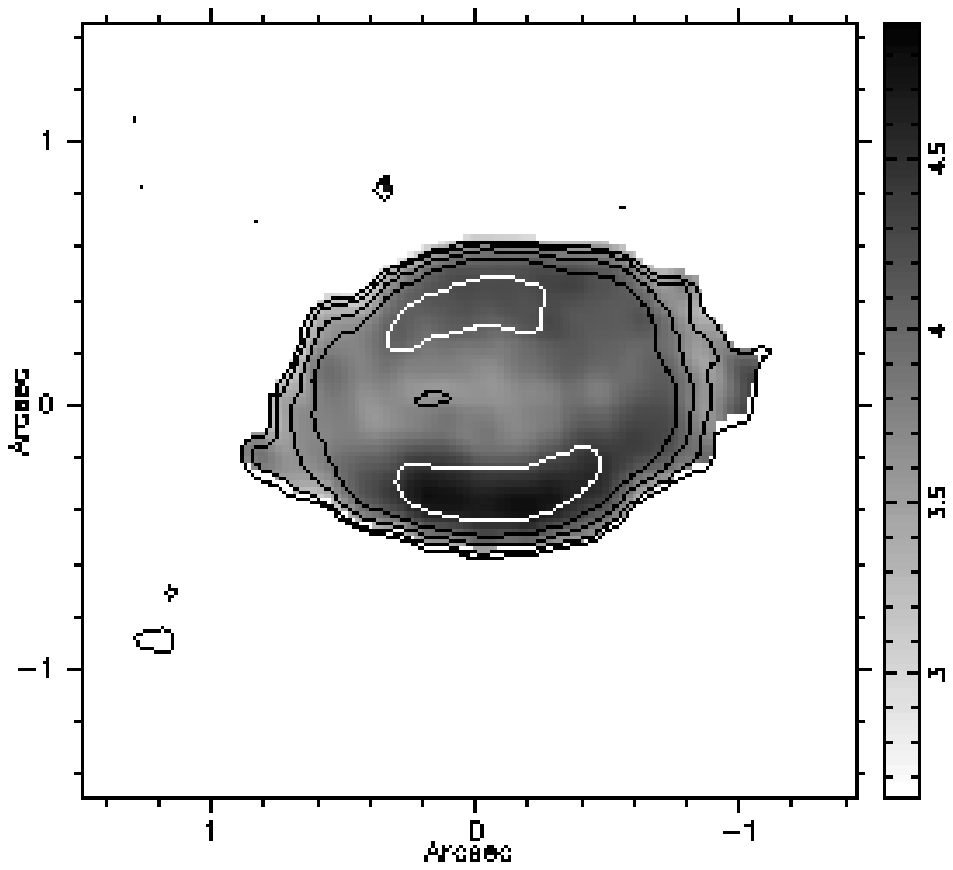}
\end{center}
\end{figure}

\begin{figure}
\begin{center}
\includegraphics[width=80mm]{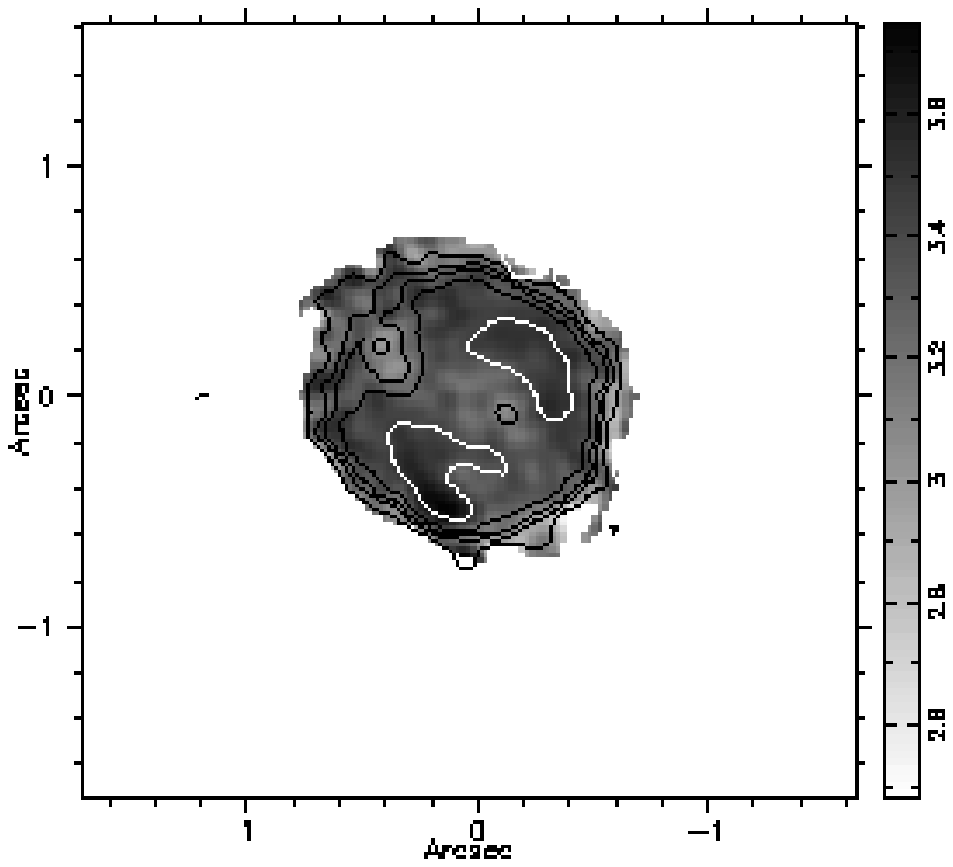}
\end{center}
\end{figure}

\clearpage

\clearpage

\clearpage

\clearpage
\begin{table}
\begin{center}
\caption{Basic Information for five compact PNs}
\label{tb:objects-dust}
\begin{tabular}{ccccc} \hline \hline
  Object & PN G & RA & Dec & Radio Diameter ($''$) \\ \hline
  BD+30$^\circ$3639 & $064.7+05.0$ & 19:34:45.2 & $+$30:30:59 & 6 \\
  IC~5117 & $089.8-05.1$ & 21:32:31.0 & $+$44:35:48 & 1.5 \\ 
  M~1-61 &  $019.4-05.3 $ & 18:45:56.0 & $-$14:27:42 & 1.8 \\
  M~2-42 & $027.6+04.2$ & 18:26:40.2 & $-$02:42:58 & 1.5 \\
  M~3-35 & $071.6-02.3$ & 20:21:03.8 & $+$32:29:22 & 1.5 \\ \hline \hline
\end{tabular}
\end{center}
\end{table}

\begin{table}
\begin{center}
\caption{The angular resolutions of the radio and optical images}
\label{tb:fwhm}
\begin{tabular}{cccc} \hline \hline
  \multicolumn{1}{c}{Object}&\multicolumn{2}{c}{VLA}&\multicolumn{1}{c}{HST}\\ 
& Beam size ($''$) & Beam PA ($^\circ$) & FWHM ($''$)\\ \hline
  BD+30$^\circ$3639 & 0.082 $\times$ 0.078 & -28.37 & 0.11 \\
  IC~5117 & 0.13 $\times$ 0.12 & -85.27 & 0.085 \\ 
  M~1-61 & 0.20 $\times$ 0.14 & 15.98  & 0.085 \\
  M2~43 & 0.15 $\times$ 0.13 & 25.51 & 0.095 \\
  M~3-35 & 0.12 $\times$ 0.12 & -81.02 & 0.085 \\ \hline \hline
\end{tabular}
\end{center}
\end{table}
\begin{table}
\begin{center}
\caption{The angular distances shifted in H$\alpha$ images}
\label{tb:shift}
\begin{tabular}{cccc} \hline \hline
  Object & Eastward ($''$) & Northward ($''$) & Offset Distance ($''$)\\ \hline
  BD+30$^\circ$3639 & $-0.364$ & $-0.455$ & 0.583 \\
  IC~5117 & $-0.536$ & $-0.378$ & 0.656 \\ 
  M~1-61 &  $-0.261$ & $+0.261$ & 0.369 \\
  M~2-43 & $+1.548$ & $-0.273$ & 1.572 \\
  M~3-35 & $-1.138$ & $-0.299$ & 1.177 \\ \hline \hline
\end{tabular}
\end{center}
\end{table}

\begin{table}
\begin{center}
\label{tb:espar}
\caption[DES model parameters of PNs]
	{DES model parameters of planetary nebulae}
\begin{tabular}{c|c c c c c c c c} \hline \hline
        Object & $a$ & $b$ & $\Delta b$ & $\beta$ & $\alpha$ & $i$ &
           $\delta_{\phi}$ & $\delta_{\theta}$ \\ \hline
        BD+30$^{\circ}$3639  & 18 & 15 & 5 & 0.1 & 1 & 15 & 0.9 & 1  \\
        IC5117 & 20 & 12 & 5 & 0.1 & 1 & 40 & 1 & 1  \\
        M1-61 & 12 & 10 & 8 & 0.2 & 1 & 70 & 0.9 & 1  \\
	M2-43 & 20 & 12 & 6 & 0.3 & 1 & 45 & 0.7 & 1 \\
        M3-35 & 10 & 12 & 10 & 0.4 & 1 & 80 & 1 & 1  \\ \hline \hline
\end{tabular}
\end{center}
\end{table}

\end{document}